\documentclass[seceq]{ptptex}
\usepackage{amsmath,amssymb,amsfonts}
\usepackage{graphicx}
\usepackage{color}

\newcommand{\Comments}[1]{}

\newcommand{\be}{\begin{equation}}
\newcommand{\ee}{\end{equation}}
\newcommand{\ba}{\begin{eqnarray}}
\newcommand{\ea}{\end{eqnarray}}
\newcommand \nn {\nonumber}

\newcommand{\VEV}[1]{\langle{#1}\rangle}
\newcommand{\chibar}{{\bar{\chi}}}
\newcommand{\Psfig}[2]{\includegraphics[width=#1]{Figs/#2}}
\newcommand{\Feff}{{\cal F}_\mathrm{eff}}
\newcommand{\Vq}{{\cal V}_{q}}
\newcommand{\Fq}{{\cal F}_{q}}

\newcommand{\Od}{{\cal O}}
\newcommand{\Vp}{V^{+}}
\newcommand{\Vm}{V^{-}}
\newcommand{\bsig}{b_\sigma}
\newcommand{\bsigp}{b'_\sigma}
\newcommand{\bsigt}{\tilde{b}_\sigma}

\newcommand{\mq}{\tilde{m}_q}

\newcommand{\Fint}{{\cal D}}
\newcommand{\Seff}[2]{S_\mathrm{#1}^{(#2)}}
\newcommand{\Disp}[1]{{\displaystyle{#1}}}

\newcommand{\Zchi}{Z_{\chi}}
\newcommand{\Zp}{Z_{+}}
\newcommand{\Zm}{Z_{-}}
\newcommand{\Zpm}{Z_{\pm}}
\newcommand{\tilmu}{\tilde{\mu}}
\newcommand{\phiT}{\varphi_\tau}

\newcommand{\psiT}{\psi_{\tau}}
\newcommand{\psiS}{\psi_s}
\newcommand{\psiTT}{\psi_{\tau\tau}}
\newcommand{\psiTS}{\psi_{\tau s}}
\newcommand{\psiSS}{\psi_{ss}}

\newcommand{\psibarT}{\bar{\psi}_{\tau}}
\newcommand{\psibarS}{\bar{\psi}_s}
\newcommand{\psibarTT}{\bar{\psi}_{\tau\tau}}
\newcommand{\psibarTS}{\bar{\psi}_{\tau s}}
\newcommand{\psibarSS}{\bar{\psi}_{ss}}

\newcommand{\wt}{\omega_\tau}

\newcommand{\bt}{\beta_\tau}
\newcommand{\bs}{\beta_s}
\newcommand{\bts}{\beta_{\tau s}}
\newcommand{\btt}{\beta_{\tau\tau}}
\newcommand{\bss}{\beta_{ss}}

\notypesetlogo                       

\title{
Effective Potential in the Strong-coupling Lattice QCD\\
with Next-to-Next-to-Leading Order Effects
}
\author{
Takashi Z. \textsc{Nakano}$^1$, 
Kohtaroh \textsc{Miura}$^2$
and
Akira \textsc{Ohnishi}$^2$
}
\inst{
$^1$Department of Physics, Kyoto University, Kyoto 606-8502, Japan\\
$^2$Yukawa Institute for Theoretical Physics, Kyoto University,
Kyoto 606-8502, Japan
}

\abst{
We derive an analytic expression of the effective potential
at finite temperature ($T$) and chemical potential ($\mu$)
in the strong-coupling lattice QCD for color $SU(3)$ including 
next-to-next-to-leading order (NNLO) effects in the strong coupling expansion.
NNLO effective action terms are systematically evaluated
in the leading order of the large dimensional ($1/d$) expansion,
and are found to come from some types of connected two plaquette configurations.
We apply the extended Hubbard-Stratonovich transformation
and a gluonic dressed fermion technique
to the effective action,
and obtain the effective potential as a function of
$T$, $\mu$, and two order parameters;
chiral condensate and a vector potential field.
The next-to-leading order (NLO) and NNLO effects
result in modifications of
the wave function renormalization factor, 
quark mass and chemical potential.
We find that $T_{c,\mu=0}$ and $\mu_{c,T=0}$ are similar to the NLO results,
whereas the position of the critical point is sensitive to NNLO corrections.
}
\begin{document}
\maketitle

\section{Introduction}
\label{sec:Intro}

Understanding the Quantum Chromodynamics (QCD) phase diagram is
one of the most interesting problems in quark and hadron physics.
In the present Relativistic Heavy-Ion Collider (RHIC) experiments,
phase transition to strongly coupled matter consisting of quarks and gluons
seems to be observed.\cite{Muller:2006ee}
The QCD phase transition observed at RHIC takes place
at almost zero baryon density, 
where the predictions of lattice QCD Monte-Carlo (MC) simulations are reliable.
The phase transition of compressed baryonic matter will be probed
in the future experiments in FAIR, J-PARC and low energy programs at RHIC.
At finite baryon densities, the lattice MC simulations are difficult
due to the complex fermion determinant.\cite{MC-sign-problem}
In order to discuss the whole shape of the phase diagram,
it is necessary to invoke some approximations in QCD
or to apply effective models.
The strong-coupling lattice QCD (SC-LQCD) is
one of the most instructive approximations to investigate
the phase structure at finite temperature $T$ and chemical potential $\mu$.

SC-LQCD was first applied to the pure Yang-Mills theory.
Wilson showed that the Wilson loop would follow the area law
at strong coupling,\cite{Wilson:1974sk}
and the Creutz demonstrated that the lattice MC simulation can 
connect strong coupling and weak coupling (perturbative) expressions
of the string tension.\cite{Creutz}
The behavior of the string tension as a function of the inverse coupling,
$\beta=2N_c/g^2$, is well understood in the strong coupling and character
expansions, which were developed by M\"unster~\cite{Munster:1980iv}.
Chiral symmetry in SC-LQCD also has been long studied from 1980s.
Basic formulations has been developed 
based on the staggered~\cite{Kawamoto:1981hw,Hoek:1981uv}, 
Wilson~\cite{Smit:1980nf,Kawamoto:1981hw}
and na\"ive~\cite{KlubergStern:1981wz} fermions.
The domain-wall \cite{Kaplan:1992bt} 
and the overlap \cite{Neuberger:1998wv} fermion
provide modern formulation of the lattice chiral symmetry,
and some SC-LQCD based investigations are found in Refs.
\citen{Brower:1999ak,Levkova:2004xw} (domain-wall) and
\citen{IchinoseNagao,XQLuo} (overlap).
In the strong coupling limit (SCL),
chiral symmetry is spontaneously broken in vacuum
\cite{Kawamoto:1981hw,Hoek:1981uv,Smit:1980nf,KlubergStern:1981wz,Brower:1999ak,Levkova:2004xw,IchinoseNagao},
and restored at high $T$ and/or large $\mu$~\cite{XQLuo,Damgaard,DHK1985,Faldt:1985ec,Ilgenfritz:1984ff,Gocksch:1984yk,Bilic,Bilic:1991nv,SCL_PLoop,Nishida:2003uj,Fukushima:2003vi,Nishida:2003fb,Azcoiti:2003eb,Kawamoto:2005mq}.

Since the SC-LQCD is based on the same formulation
as the lattice MC simulations,
its results should be consistent with the MC results such as 
hadron mass spectrum~\cite{Kawamoto-Shigemoto-hadron-mass,KlubergStern:1982bs,Jolicoeur:1983tz} 
as long as the applied approximations are valid.
The phase diagram structure has been predicted in the strong coupling 
limit,\cite{Damgaard,Bilic:1991nv,Nishida:2003uj,Fukushima:2003vi,Nishida:2003fb,Kawamoto:2005mq}
and it is recently confirmed qualitatively in MC simulations
\cite{deForcrand:2009dh}
based on the monomer-dimer-polymer (MDP) formalism~\cite{Karsch:1988zx}.
In order to make a step forward towards the true phase diagram,
it is necessary to develop the formalism to include the plaquette effects
both in lattice MC simulations at finite $\mu$ and SC-LQCD.
There exists some SC-LQCD works
including NLO effects, {\em i.e.} one plaquette contributions,
on the hadron masses~\cite{Kawamoto-Shigemoto-hadron-mass,KlubergStern:1982bs,Jolicoeur:1983tz}
and the phase diagram~\cite{Bilic,PDevol,Miura:2008gd}.
In our previous work on the NLO SC-LQCD,\cite{PDevol}
we find that the phase diagram evolves to an empirical shape
with increasing $\beta=6/g^2$,
while the critical temperature at zero chemical potential $T_{c,\mu=0}$
is larger than the MC results.
The later observation suggests that we need to evaluate 
the next-to-next-to-leading order (NNLO) effects on the phase diagram,
which have never been investigated before.

In this paper, we derive an analytic expression of the effective potential
including NNLO effects at finite $T$ and $\mu$,
and investigate NNLO contributions to the phase diagram.
We adopt one species of unrooted staggered fermion corresponding to $N_f=4$
in the continuum region.
Since the flavor dependence of the phase boundary has been shown to be
moderate~\cite{Bilic:1991nv,Nishida:2003fb,D'Elia:2002gd},
the present results could be valuable for the understanding
of the phase diagram with $N_f=2+1$.
Effective action terms from one and two plaquette configurations
are obtained by integrating out spatial link variables.
We apply the extended Hubbard-Stratonovich
transformation~\cite{PDevol,Miura:2008gd}
to bosonize fermion interaction terms.
With the $1/g^4$ corrections,
we encounter those terms containing  
the next-to-nearest neighbor (NNN) interaction,
which can be evaluated by introducing a gluonic dressed fermion.
The effective potential is obtained as a function of
$T$, $\mu$ and two order parameters:
the chiral condensate and the quark number density.
We determine the equilibrium from the stationary condition
of the effective potential with respect to the auxiliary fields,
and study the properties of QCD phase diagram.
MC studies based on one species of unrooted staggered fermions
have been carried out extensively around
$\beta \sim 5$.~\cite{D'Elia:2002gd,Fodor:2001au,Forcrand,Gottlieb:1987eg,Gavai:1990ny}
For the comparison between SC-LQCD results and those in the MC simulations,
we discuss the results in the region $\beta \leq 6$.
As shown later, the comparison of NNLO and NLO results suggests
that $\beta$ values under consideration are in conversion radius.
We compare the critical temperature at $\mu=0$ ($T_{c,\mu=0}$)
with the NLO and MC results.
We also study the evolution of the phase diagram and the critical point,
as well as the possibility to have partially chiral restored matter,
which has been suggested in NLO SC-LQCD~\cite{PDevol,Miura:2008gd}.

The QCD phase transition has another aspect of the deconfinement transition.
In order to discuss the deconfinement transition,
it is necessary to include the Polyakov loop effects
as discussed 
in the framework of SC-LQCD\cite{Ilgenfritz:1984ff,Gocksch:1984yk,SCL_PLoop}
and in the Nambu-Jona-Lasinio model with the Polyakov loop (PNJL)
~\cite{Fukushima:2003fw,Ratti:2005jh}.
In this study, we concentrate on the chiral phase transition,
and the simultaneous description of chiral and deconfinement transitions
will be reported elsewhere.\cite{NNLOwithPolyakov}

On the phase transition study with unrooted staggered fermion,
we still have some debate.
In the strong coupling limit, the effective action with one species of unrooted staggered fermion
is in the same universality class as the three-dimensional
$O(2)$ spin models, and the phase transition is the second order
.\cite{Boyd:1991fb}
In the continuum region, one species of unrooted staggered fermion corresponds
to four flavors (or tastes).
The phase transition at $\mu=0$ is expected to be the first order
for $N_f \geq 3$ from the anomaly argument.\cite{Pisarski:1983ms}
The axial anomaly may cancel from the doublers on the lattice,\cite{Creutz:2007yg}
but it is discussed that the conventional staggered anomaly appears
in the taste-singlet PCAC relation.\cite{Kronfeld:2007ek}
In numerical MC simulations,\cite{D'Elia:2002gd}
the transition at $\mu=0$ is shown
to be the first order in the continuum region.
While we have these complications,
the staggered fermions has a merit that it is simple.
As a result, it is fast in numerical simulations,
and it is possible to carry out the analytic calculations in the mean field
approximations when we adopt the unrooted staggered fermion.

This paper is organized as follows.
In \S \ref{Sec:Feff}, we derive the effective potential
with $1/g^4$ corrections. 
In \S \ref{Sec:Results}, we show the calculated results
of the effective potential and the phase structure.
We summarize our work in \S \ref{Sec:Summary}. 
All through this paper, we use the lattice unit $a=1$,
and physical values are shown in dimensionless values 
normalized by the lattice spacing $a$.

\section{Effective Potential in NNLO SC-LQCD}
\label{Sec:Feff}

In this section,
we derive the effective potential in SC-LQCD
with NNLO corrections ($1/g^4$)
at finite temperature $T$ and chemical potential $\mu$.
We start from the lattice QCD action 
with one species of unrooted staggered fermion ($n_f=1$)
for color $\mathit{SU}(N_c)$~(\S~\ref{subsec:S_LQCD}),
which corresponds to $N_f=4n_f=4$ in the continuum limit.
After a short review on SCL and NLO effective action~(\S~\ref{subsec:SCL-NLO}),
we derive the NNLO effective action (\S~\ref{subsec:S_eff}).
Interaction terms are reduced into a bilinear form of fermions
through the extended Hubbard-Stratonovich transformation (\S~\ref{subsec:EHS})
and by introducing a gluonic dressed fermion~(\S~\ref{subsec:dressed_fermion}).
We obtain an analytic expression of the effective potential
in \S~\ref{subsec:Feff}.

\subsection{Lattice QCD action}
\label{subsec:S_LQCD}

The lattice QCD action and the partition function
with one species of staggered fermion
for color $\mathit{SU}(N_c)$ are given as follows
\begin{align}
\label{Eq:ZLQCD}
&{\cal Z}_{\mathrm{LQCD}} 
= \int \Fint[\chi,\chibar,U_\nu]~e^{-S_\mathrm{LQCD}}
= \int \Fint[\chi,\chibar,U_\nu]~e^{-S_F^{(\tau)}-S_F^{(s)}-S_G}
\ ,\\
&S_F^{(\tau)} = \frac {1}{2} \sum_x (V_x^+ - V_x^-)+ m_0 \sum_x M_x
\ ,\\
&S_F^{(s)}
 =\frac {1}{2} \sum_x \sum_{j=1}^d \eta_{j,x} 
  \left[ \bar{\chi}_x U_{j,x} \chi_{x + \hat{j}} - \bar{\chi}_{x + \hat{j}}
   U_{j,x}^{\dagger} \chi_x \right]
\ ,\\
&S_G
 =
 - \frac {1}{g^2} \sum_x \sum_{j>0} \left[
 	\mathrm{tr}\,U_{0j,x} + \mathrm{tr}\,U_{0j,x}^{\dagger}
 	\right]
 - \frac{1}{g^2}\sum_x \sum_{k>j>0} \left[
	  \mathrm{tr}\,U_{jk,x}
	+ \mathrm{tr}\,U_{jk,x}^{\dagger} \right]
\ ,
\end{align}
where
$\chi (\bar{\chi})$, $m_0$, $U_{0,x}(U_{j,x})$,
and
$U_{0j,x}$ ($U_{jk,x}$)
denote
the quark (anti-quark) field,
the bare quark mass, temporal (spatial) link variable,
and temporal (spatial) plaquette, respectively.
The spinor structure is compressed into
the staggered phase factor 
$\eta_{j,x}=(-1)^{x_0+\cdots +x_{j-1}}$~\cite{Susskind:1976jm,Sharatchandra:1981si,Kawamoto:1981hw}.
We have introduced two types of mesonic composites
\begin{align}
V_x^+ =& \bar{\chi}_x e^{\mu} U_{0,x} \chi_{x + \hat{0}}
\ ,\quad
V_x^- = \chibar_{x+\hat{0}} e^{-\mu} U_{0,x}^{\dagger} \chi_x
\ ,\quad
M_x = \chibar_x \chi_x
\ ,
\end{align}
which appear in the effective action discussed later.
Quark chemical potential $\mu$ on the lattice
is introduced as a weight of the temporal hopping
in $V^{\pm}$~\cite{Hasenfratz:1983ba}.
By using a $\gamma_5$-related factor $\epsilon_x=(-1)^{x_0+\cdots+x_{d}}$,
a staggered chiral transformation is given as
$\chi_x \to e^{i\theta\epsilon_x}\chi_x$
\cite{Susskind:1976jm,Sharatchandra:1981si,Kawamoto:1981hw}.
The lattice kinetic action $S_F^{(\tau,s)}$
is invariant under this chiral transformation
in the chiral limit $m_0\to 0$.
Throughout this paper, we consider the case of color $\mathit{SU}(N_c=3)$
in 3+1 dimension $(d=3)$ spacetime.
Temporal and spatial lattice sizes are denoted as $N_\tau$ and $L$,
respectively.
While $T=1/N_{\tau}$ takes discrete values, 
we consider $T$ as a continuous valued temperature.
We take account of finite $T$ effects
by imposing periodic and anti-periodic 
boundary conditions on link variables and quark fields, respectively.
We take the static and diagonalized gauge (called Polyakov gauge)
for temporal link variables with respect for the
periodicity~\cite{Damgaard}.
In these setups, we evaluate the effective potential
$\Feff=-\log\bigl[{\cal Z}_{\mathrm{LQCD}}\bigr]/(N_{\tau}L^d)$
based on the strong coupling expansion.

\subsection{Strong coupling expansion and SCL and NLO effective action}
\label{subsec:SCL-NLO}
In a finite $T$ treatment of SC-LQCD,
we first derive an effective action
by integrating out the spatial links $U_j$,
and the temporal link variable $U_0$ is evaluated later.
The $U_j$ integral can be exactly performed
in each order of the $1/g^2$ expansion
by utilizing the link integral formulae
\begin{align}
\label{Eq:OneLinkA}
&\int dU\, U_{ab}\, U^\dagger_{cd}
=\frac{1}{N_c} \delta_{ad}\,\delta_{bc}
\ ,\qquad
\int dU\, U_{ab}\, U_{cd}\, \cdots U_{ef} 
= \frac{1}{N_c!} \varepsilon_{ac\cdots e}\,\varepsilon_{bd\cdots f}
\ ,
\end{align}
and so on.
Then an isotropic hopping structure of hadronic composites emerges,
and enables us to simplify the effective action
by utilizing another expansion: $1/d$ expansion,
which is explained in the next subsection.

In this work, we consider the effective action
including the SCL ($1/g^0$), NLO ($1/g^2$) and NNLO ($1/g^4$) terms
in the strong coupling expansion,
while we keep only the leading order terms ($1/d^0$) in the $1/d$ expansion.
The effective action is defined as
\begin{align}
e^{-S_\mathrm{eff}(\chi,\chibar,U_0)}
=& \int \mathcal{D} U_j~e^{-S_\mathrm{LQCD}}
= e^{-S_F^{(\tau)}} \int \mathcal{D}U_j~e^{-S_F^{(s)}-S_G}
\nonumber\\
=& e^{-S_F^{(\tau)}-S_\mathrm{SCL}^{(s)}}
\ \big\langle e^{-S_G} \big\rangle
\ .\label{Eq:Seff}
\end{align}
We have defined an expectation value as
\begin{align}
\big\langle\mathcal{O}\big\rangle
=&\frac{1}{Z_{\mathrm{SCL}}^{(s)}}
\int\mathcal{D}U_j~\mathcal{O}[U_j]~e^{-S_{F}^{(s)}}
\ ,\quad
Z_{\mathrm{SCL}}^{(s)}
=\int\mathcal{D}U_j~e^{-S_{F}^{(s)}}
=e^{-S_\mathrm{SCL}^{(s)}}
\ .
\end{align}
The factor $1/Z_\mathrm{SCL}^{(s)}$
ensures the normalization property $\langle \mathbf{1}\rangle =1$.
In order to systematically evaluate
the effective action terms in each order of $1/g^2$,
the cumulant (or coupled cluster) expansion is indispensable.
It is well known that the expectation value of the exponential form operator 
with a small factor ({\em i.e.} $1/g^2$) can be evaluated
by using the cumulant expansion~\cite{cumulant}
\begin{align}
\big\langle e^{-S_G} \big\rangle
=\sum_{n=0}^\infty \frac{(-1)^n}{n!} \big\langle{S_G}^n\big\rangle
=\exp\biggl[\sum_{n=1}^{\infty}
\frac{(-1)^n}{n!}
\big\langle S_G^n \big\rangle_c
\biggr]
\ .\label{Eq:cum}
\end{align}
The correlation part in the connected diagram contributions
is shown by the bracket $\langle\cdots\rangle_c$, and  is called a cumulant,
e.g. $\VEV{S_G^2}_c=\VEV{S_G^2}-\VEV{S_G}^2$.
By substituting Eq.~(\ref{Eq:cum}) into Eq.~(\ref{Eq:Seff}),
we find that the effective action is obtained as
\begin{align}
S_\mathrm{eff}
=&S_\mathrm{SCL}-\sum_{n=1}^{\infty}
\frac{(-1)^n}{n!}
\big\langle S_G^n \big\rangle_c
\nonumber\\
=&S_\mathrm{SCL}+\Delta S_\mathrm{NLO}+\Delta S_\mathrm{NNLO}
+\Od(1/g^6, 1/\sqrt{d})
\ .\label{Eq:SeffB}
\end{align}
The $n$-th term in the sum is proportional to $1/g^{2n}$,
and we can identify
$n=1$ and $n=2$ terms as NLO and NNLO effective action,
$\Delta S_\mathrm{NLO}$ and $\Delta S_\mathrm{NNLO}$,
and $S_\mathrm{SCL}$ shows the SCL effective action.

\begin{figure}[bt]
\begin{center}
\Psfig{7cm}{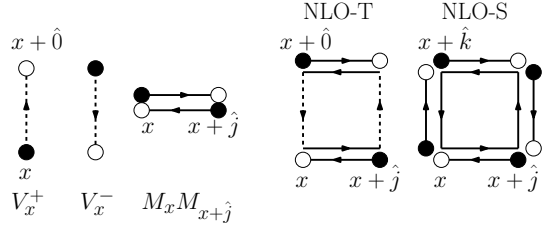}
\caption{
Diagrams contributing to the SCL and NLO effective actions.
Quarks (anti-quarks) are shown in open (filled) circles,
and spatial (temporal) link variables are 
represented by solid (dashed) lines.
Note that we should also take account of
hermite conjugate contributions for NLO diagrams.
}\label{Fig:NLOdiagram}
\end{center}
\end{figure}

The SCL and NLO effective actions are known as,\cite{Faldt:1985ec,PDevol}
\begin{align}
S_\mathrm{SCL}=S_F^{(\tau)}+S_F^{(s)}=&
S_F^{(\tau)}-\frac{1}{4N_c}\sum_{x,j>0}M_xM_{x+\hat{j}}
\label{Eq:SeffSCL}
\ ,\\
\Delta S_\mathrm{NLO}
= \langle S_G \rangle_c
=& \frac {1}{4 N_c^2 g^2} \sum_{x, j>0} 
  \left[ 
   V_x^{+} V_{x + \hat{j}}^{-} + V_{x+\hat{j}}^{+} V_x^{-}
  \right]\nn\\
 -& 2\times\frac{1}{16 N_c^4 g^2} \sum_{x, k>j>0}
   M_xM_{x+\hat{j}}M_{x+\hat{k}+\hat{j}}M_{x+\hat{k}}
\ .
\label{Eq:SeffNLO}
\end{align}
The factor and sign of each term is summarized
in Table \ref{Table:factor-and-sign}.
The factor ``$2$'' in the last line 
accounts for the hermite conjugate contribution.
In Fig.~\ref{Fig:NLOdiagram},
we show the diagrams contributing to the SCL and NLO effective actions.
The first three diagrams represent the temporal hopping ($S_F^{(\tau)}$) terms
and the spatial meson hopping ($MM$) terms in $S_{\mathrm{SCL}}$.
The fourth and fifth diagrams show
$VV$ and $MMMM$ terms in $\Delta S_{\mathrm{NLO}}$.

\subsection{NNLO effective action}
\label{subsec:S_eff}

\begin{figure}[bth]
\begin{center}
\Psfig{7cm}{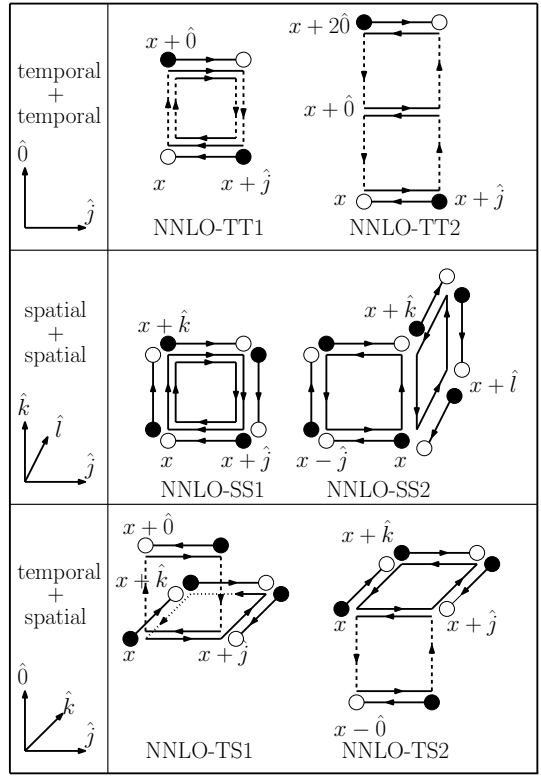}
\caption{
Diagrams contributing to the NNLO effective action.
Definitions of the symbols and lines are the same as those in 
Fig.~\protect\ref{Fig:NLOdiagram}.
The top, middle and bottom rows represent the NNLO diagrams
composed of the temporal-temporal, spatial-spatial,
and temporal-spatial plaquette configurations, respectively.
}\label{Fig:NNLOdiagram}
\end{center}
\end{figure}

From Eq.~(\ref{Eq:SeffB}), 
we find that the NNLO effective action is given as the cumulant of $S_G^2$.
\begin{align}
\Delta S_\mathrm{NNLO}
=&
-\frac{1}{2}\big\langle S_G^2\big\rangle_c
=-\frac{1}{2}\left[
\langle S_G^2\rangle
-\langle S_G\rangle^2\right]
\nonumber\\
=&-\frac{1}{2g^4}
\sum_{P,P'}
\left[
\langle U_P U_{P'} \rangle
-\langle U_P \rangle \langle U_{P'} \rangle
\right]
\ ,\label{Eq:cumNNLO}
\end{align}
where $U_P=\mathrm{tr}U_{\mu\nu,x}$ denotes the trace in the color space
of a plaquette, or its conjugate.
In the case where the two plaquettes ($P$ and $P'$) do not have
any common spatial link variables,
the average of the product is factorized as
$\VEV{U_P U_{P'}}=\VEV{U_P}\VEV{U_{P'}}$,
which cancels with the second term.
Thus only the connected diagrams can contribute to the NNLO effective action.
In Fig.~\ref{Fig:NNLOdiagram}, we show the two plaquette configurations,
where the plaquettes share at least one spatial link.

For a given two plaquette configuration,
we consider the leading order terms in the $1/d$
expansion.\cite{KlubergStern:1982bs}
The sum over spatial directions $\sum_j$ in Eq.~(\ref{Eq:SeffSCL})
would give rise to a factor $d$ due to the spatial isotropy.
Provided that the meson hopping term $\sum_jM_xM_{x+\hat{j}}$
in the SCL effective action stays finite at large $d$,
the quark field ($\chi,\bar{\chi}$) should scale as $d^{-1/4}$.
Then a factor $d$ from the sum over spatial directions
and a factor $1/d$ from the four quark fields cancel,
and the mesonic hopping term $\sum_jM_xM_{x+\hat{j}}$ is found to be
$\mathcal{O}(1/d^0)$, which is the leading order in the $1/d$ expansion.
This also applies to the NNLO diagrams.
For example, we consider the product of the same temporal plaquette,
NNLO-TT1 in Fig.~\ref{Fig:NNLOdiagram}:
We will have a factor $d$ for the sum over the temporal plaquette $U_{0j,x}$,
and four quark fields give rise to a factor $1/d$.
Thus the diagram NNLO-TT1 in Fig.~\ref{Fig:NNLOdiagram}
is in the leading order $\Od(1/d^0)$ in the $1/d$ expansion.
Diagrams with more quarks for the same plaquette configuration
are suppressed as ${\cal O}(1/\sqrt{d})$ for $N_c \geq 3$.
This is called the systematic large dimensional or $1/d$ expansion.

Since we concentrate on the leading order terms in the $1/d$ expansion,
we find that it is sufficient to consider
the diagrams shown in Fig.~\ref{Fig:NNLOdiagram},
each of which includes the minimal number of quark fields
for each plaquette configuration.
We omit two plaquette configurations
with $U_P=U_{P'}^\dagger$, which lead to a constant in the effective action
in the leading order of the $1/d$ expansion, $\Od(1/d^0)$.
By utilizing the group integral formulae Eq.~(\ref{Eq:OneLinkA}),
these contributions lead to the following effective action terms
\begin{align}
 \Delta\Seff{NNLO}{\tau\tau}
 =&
  \frac {1}{8 N_c^2 g^4} 
     \sum_{x,j>0} 
     \left[ 
	  V_x^+ V_{x+ \hat{j}}^-
	+ V_{x+\hat{j}}^+ V_x^-
     \right] \nonumber\\
  -& \frac {1}{4 N_c^3 g^4} \sum_{x,j>0} 
     \left[ 
        W_x^+W_{x+ \hat{j}}^-
      + W_{x+\hat{j}}^+W_x^- \right]
\label{tau-1/g4}
\ ,\\
\Delta\Seff{NNLO}{ss}
 =&
 - 2\times\frac {1}{32N_c^4g^4} \sum_{x,k>j>0}  
     \left[ 
      M_x M_{x + \hat{j}} M_{x + \hat{j} + \hat{k}} M_{x + \hat{k}} 
     \right]
\nn\\
 -& \frac {1}{64 N_c^7 g^4} 
     \sum_{k>0, \ \mid j \mid \neq k, \ \mid l \mid \neq \mid j \mid ,k} 
     M_x M_{x+\hat{l}}M_{x+\hat{l}+\hat{k}}M_{x +\hat{k}}
     M_{x + \hat{k}-\hat{j}} M_{x-\hat{j}}
\label{s-1/g4}
\ ,\\
\Delta\Seff{NNLO}{\tau s}
 =& \displaystyle \frac {1}{16 N_c^5 g^4} \sum_{x, j > 0, \mid k \mid
      \neq j}
     \Biggl[ 
      V_x^+ V_{x + \hat{j}}^- + V_{x + \hat{j}}^+ V_{x}^-
      + V_{x - \hat{0} + \hat{j}}^+ V_{x - \hat{0}}^- + V_{x - \hat{0}}^+
      V_{x - \hat{0} + \hat{j}}^- 
     \Biggr]
\nn\\
 &\times M_{x + \hat{j} + \hat{k}} M_{x + \hat{k}}
\label{tau-s-1/g4}
\ .
\end{align}
The factor ``$2$'' in the first line of $\Delta\Seff{NNLO}{ss}$
accounts for the hermite conjugate contribution.
We find that two new mesonic composites appear
\begin{align}
W_{x}^+ = \bar{\chi}_x e^{2 \mu} U_{0,x} U_{0, x + \hat{0}} \chi_{x + 2
 \hat{0}}
 \ ,\quad
W_{x}^- = \bar{\chi}_{x+ 2 \hat{0}} e^{-2 \mu} U_{0, x + \hat{0}}^{\dagger} U_{0,x}^{\dagger} \chi_x
\ .
\end{align}
These composite connect the quark fields in the next-to-nearest neighboring
(NNN) temporal sites.
In Fig.~\ref{Fig:NNLOdiagram},
the top, second and bottom rows represent
$\Delta\Seff{NNLO}{\tau\tau}$,
$\Delta\Seff{NNLO}{ss}$
and $\Delta\Seff{NNLO}{\tau s}$, respectively. 
Note that for diagrams which include quarks,
we also consider the hermite conjugate of those diagrams.

\begin{table}[tb]
\caption{
The factors and sign
which come from,
(A) a global NNLO factor $1/g^0$, $1/g^2$ or $1/g^4$,
(B) a symmetric factor (Sym.) $1/2$ for the product of two same plaquettes,
(C) product of factors $-1/2$ and $1/2$
    for the forward ($\chibar U\chi$) and backward
    ($\chibar U^\dagger\chi$) fermionic elements (F.E.),
(D) a product of the staggered factors $\eta_{j,x}$,
and
(E) one-link integrals.
Each entry in the left column
represents a diagram shown 
in Fig.~\protect\ref{Fig:NLOdiagram} and \protect\ref{Fig:NNLOdiagram}.
}\label{Table:factor-and-sign}
\renewcommand{\arraystretch}{1.4}
\begin{center}
\begin{tabular}{l|ccccc|r}
\hline
          &(A)~Global	&(B)~Sym.	&(C)~F.E.	&(D)~$\eta_{j,x}$	&(E)~$U_j$	& Total \\\hline
SCL-S     &$1/g^0$ 	&1   		&$-(1/2)^2$ 	&$+1$ 			&$1/N_c$ 	&$-1/4 N_c$\\\hline
NLO-T     &$1/g^2$ 	&1   		&$-(1/2)^2$ 	&$-1$ 			&$1/N_c^2$ 	&$+1/4 N_c^2g^2$\\
NLO-S     &$1/g^2$ 	&1   		&$+(1/2)^4$ 	&$-1$ 			&$1/N_c^4$ 	&$-1/16N_c^4g^2$\\\hline
NNLO-TT1  &$1/g^4$ 	&1/2 		&$-(1/2)^2$ 	&$-1$ 			&$1/N_c^2$ 	&$+1/8 N_c^2g^4$\\
NNLO-TT2  &$1/g^4$ 	&1   		&$-(1/2)^2$ 	&$+1$ 			&$1/N_c^3$ 	&$-1/4 N_c^3g^4$\\
NNLO-SS1  &$1/g^4$ 	&1/2 		&$+(1/2)^4$ 	&$-1$ 			&$1/N_c^4$ 	&$-1/32N_c^4g^4$\\
NNLO-SS2  &$1/g^4$ 	&1   		&$-(1/2)^6$ 	&$+1$ 			&$1/N_c^7$ 	&$-1/64N_c^7g^4$\\
NNLO-TS1,2&$1/g^4$ 	&1   		&$+(1/2)^4$ 	&$+1$ 			&$1/N_c^5$ 	&$+1/16N_c^5g^4$\\\hline
\end{tabular}
\end{center}
\renewcommand{\arraystretch}{1.0}
\end{table}

We summarize the factor and sign
in SCL, NLO and NNLO in Table~\ref{Table:factor-and-sign}.
The factor and sign of each term is a product of
(A) a global NNLO factor $1/g^0$, $1/g^2$ or $1/g^4$,
(B) a symmetric factor $1/2$ for the product of two same plaquettes,
(C) product of factors $-1/2$ and $1/2$
    for the forward ($\chibar U\chi$) and backward
    ($\chibar U^\dagger\chi$) fermionic elements,
(D) product of the staggered factors $\eta_{j,x}$,
and
(E) a factor from the one-link integral.
Expressing
the effective action as a function of 
color singlet composites $M,~V$ and $W$,
we generally obtain a factor ``$-1$''
resulting from an odd number times of
grassmann number exchanges.
This sign factor is taken care in the global factor
and shown in the column (A)
of Table~\ref{Table:factor-and-sign}.
For a product of different plaquettes $(P,P')$,
there are two combinations $(P,P')$ and $(P',P)$
in the sum of Eq.~(\ref{Eq:cumNNLO}),
and they give rise to a factor $2$.
With the factor $-1/2g^4$ found in Eq.~(\ref{Eq:cumNNLO}),
we get a global factor $1/g^4$.
For a product of the same plaquettes $P=P'$
in NNLO-TT1 and NNLO-SS1, we need a symmetric factor $1/2$.
The factor from the one-link integral is $1/N_c^k$,
where $k$ is the number of integrated out spatial links.
Also in the case to use the second formula in Eq.~(\ref{Eq:OneLinkA}),
we have $1/N_c$ in the present configurations.
For example, in NNLO-TT1, 
we have a factor $1/(N_c!)^2$ from the spatial link integrals,
and the following combination of temporal link variables remains
on the link $(0,x+\hat{j})$
\begin{align}
\varepsilon_{a_1 b_1 \cdots c_1} \varepsilon_{a_2 b_2 \cdots c_2}
 U_{0,x+\hat{j}}^{b_1 b_2}\cdots U_{0, x+\hat{j}}^{c_1 c_2}
=(N_c-1)! \times U_{0,x+\hat{j}}^{\dagger\, a_2 a_1}
\ .
\end{align}
We also find another $(N_c-1)!$ appears on the link $(0,x)$,
and the factor from the one-link integral is found to be $1/N_c^2$.
In a similar way, $1/N_c^4$ appears for NNLO-SS1.
Note that NLO-S and NNLO-SS1 action terms are hermite,
hence one find a factor $2$ in $MMMM$ terms 
of Eq.~(\ref{Eq:SeffNLO}) and (\ref{s-1/g4}).

We explain the factor and sign for NNLO-TS1, as an example.
There are two forward ($(k,x)$ and $(j,x+\hat{k})$)
and two backward ($(k,x+\hat{j})$ and $(j,x+\hat{0})$) fermionic elements,
and we have a factor $+(1/2)^4$.
Staggered factors for these elements are
$\eta_{k,x}$,
$\eta_{j,x+\hat{k}}=\alpha\eta_{j,x}$,
$\eta_{k,x+\hat{j}}=-\alpha\eta_{k,x}$
and
$\eta_{j,x+\hat{0}}=-\eta_{j,x}$,
where $\alpha=1~(-1)$ for $k>j$ ($k<j$).
The product of these staggered factors results in $+1$.
Five spatial links are integrated out, and give a factor $1/N_c^5$.
Together with the global factor $1/g^4$, 
we find the coefficient $+1/16N_c^5g^4$ for NNLO-TS1.

We combine those terms in NLO and NNLO having
the same composites.
Now we have obtained the effective action $S_{\mathrm{eff}}$
in Eq.~(\ref{Eq:SeffB}) up to $\mathcal{O}(1/g^6,1/\sqrt{d})$
\begin{subequations}
\begin{align}
&S_\mathrm{eff}
= S_\mathrm{SCL} 
 + \Delta S^{\tau}
 + \Delta S^{s}
 + \Delta S^{\tau\tau}
 + \Delta S^{ss}
 + \Delta S^{\tau s}
\\
&= \frac12 \sum_x (V^+_x - V^-_x) - \frac{b_\sigma}{2d}\sum_{x,j>0} [MM]_{j,x}
\label{Eq:SeffSCL2}
\\
&+ \frac12\,\frac {\bt}{2d} \sum_{x, j>0} 
  [V^+V^- + V^-V^+]_{j,x}
\label{Eq:SeffT}
\\
&- \frac12\,\frac{\beta_s}{d(d-1)} \sum_{x, j>0,k>0,k\not=j} 
  [MMMM]_{jk,x}
\label{Eq:SeffS}
\\
&-\frac {\beta_{\tau\tau}}{2d} \sum_{x,j>0} 
  [W^+W^- + W^-W^+]_{j,x}
\label{Eq:SeffTT}
\\
&-\frac {\beta_{ss}}{4d(d-1)(d-2)}
     \sum_{\substack{x,\,j>0,\,|k|>0,\,|l|>0\\|k|\neq j,\,|l|\neq j,\,|l|\neq|k|}}
     [MMMM]_{jk,x}
     [MM]_{j,x+\hat{l}}
\label{Eq:SeffSS}
\\
&+ \frac {\beta_{\tau s}}{8d(d-1)} \sum_{x, j > 0, \mid k \mid \neq j}
  [V^+V^- + V^-V^+]_{j,x}
     \left( [MM]_{j,x+\hat{k}} + [MM]_{j,x+\hat{k}+\hat{0}} \right)
\label{Eq:SeffTS}
\ .
\end{align}
\end{subequations}
We have introduced a short-hand notation
\begin{align}
&[AB]_{j,x} = A_xB_{x+\hat{j}}\ ,\quad
[ABCD]_{jk,x}= A_xB_{x+\hat{j}}C_{x+\hat{j}+\hat{k}}D_{x + \hat{k}} \ .
\end{align}
We obtain Eq.~({\ref{Eq:SeffTS}}) by shifting $x$ to $x+\hat{0}$
in the third and fourth $V^+V^-$ terms in Eq.~({\ref{tau-s-1/g4}}).
NLO and NNLO effective action terms,
$\Delta S^{K} (K=\tau,s,\tau\tau,ss,\tau s)$,
correspond to Eqs.~(\ref{Eq:SeffT})--(\ref{Eq:SeffTS}).
The coefficients are defined as
\begin{align}
&b_\sigma=\frac{d}{2N_c}\ ,\quad
\bt=\frac{d}{N_c^2g^2}\,\left(1+\frac{1}{2g^2}\right)
\ ,\quad
\beta_s=\frac{d(d-1)}{8N_c^4g^2}\,\left(1+\frac{1}{2g^2}\right)
\ ,\label{Eq:CoeffA}\\
&\beta_{\tau\tau}=\frac{d}{2N_c^3g^4}
\ ,\quad
\beta_{ss}=\frac{d(d-1)(d-2)}{16N_c^7g^4}
\ ,\quad
\beta_{\tau s}=\frac{d(d-1)}{2N_c^5g^4}
\ .\label{Eq:CoeffB}
\end{align}

\subsection{Extended Hubbard-Stratonovich transformation}
\label{subsec:EHS}

The effective action $S_\mathrm{eff}$ derived in the previous subsection
contains several types of composite products, which include many fermion fields.
In order to perform the grassmann integral,
it is much more convenient to reduce these higher order terms
to the spatially local and bilinear form in the fermion fields
through the bosonization procedure,
so-called the Hubbard-Stratonovich (HS) transformation.
The standard HS transformation is applicable
to the product of composites of similar kind,
while NLO and NNLO terms contain the product of different types,
such as $V^+V^-$.
In order to treat these terms,
we apply the extended Hubbard-Stratonovich (EHS) 
transformation~\cite{PDevol,Miura:2008gd},
where two auxiliary fields $(\varphi, \phi)$ 
are introduced simultaneously
\begin{align}
e^{\alpha A B}  
&= \int\, d\varphi\, d\phi\,
	e^{-\alpha\left\{
        \left[
		 \varphi-(A+B)/2
        \right]^2
        +\left[
		 \phi - i(A-B)/2
        \right]^2
		\right\}+ \alpha AB}
\nonumber\\
&\approx 
	\left.
	e^{-\alpha\left\{
		\varphi^2-(A+B)\varphi
		+ \phi^2 - i(A-B)\phi
		\right\}}\right|_\mathrm{stationary}
\label{Eq:EHSa}
\\
&\approx
	\left. e^{-\alpha\left\{
		\bar{\psi}\psi-{A}\psi-\bar{\psi}B
		\right\}}\right|_{\mathrm{stationary}}
\label{Eq:EHSc}
\ ,
\end{align}
where 
$\alpha$ is an arbitrary positive real constant.
In the second line, we approximate the integral
by the integrand at the stationary values,
$\varphi=\VEV{A+B}/2$ and $\phi=i\VEV{A-B}/2$
(the saddle point approximation). 
This procedure is denoted by ``$\approx$'',
and we use this notation in later discussions.
In the last line Eq.~(\ref{Eq:EHSc}), 
we have performed the transformation
$\psi=\varphi+i\phi$ and $\bar{\psi}=\varphi-i\phi$.
When $A=B$ is satisfied, 
Eq.~(\ref{Eq:EHSa}) is equivalent
to the standard HS transformation.
In the case where both $\VEV{A}$ and $\VEV{B}$ are real,
the stationary value of $\phi$ is pure imaginary.
Thus we replace $\phi \rightarrow i \omega$ in Eq.~(\ref{Eq:EHSa}),
then the stationary value of $\omega$ is real
\begin{align}
e^{\alpha A B}
&\approx
	\left.
	e^{-\alpha\left\{
		\varphi^2-(A+B)\varphi
		- \omega^2 +(A-B)\omega
		\right\}}\right|_\mathrm{stationary}
\label{Eq:EHSb}
\ .
\end{align}

\begin{table}[tb]
\caption{
The application of 
the Extended Hubbard-Stratonovich transformation
to effective action terms.
}\label{Table:EHSsubst}
\begin{center}
\begin{tabular}{ll|ccc}
\hline
&&$\alpha$ & $A$ & $B$\\
\hline
$\Delta S^{\tau\tau}$ & (\protect{\ref{Eq:SeffTT}})
& $\beta_{\tau\tau}/2d$ & $W_x^+$ & $W_{x+j}^-$\\
$\Delta S^{ss}$ &(\protect{\ref{Eq:SeffSS}})
& $\beta_{ss}/4d(d-1)(d-2)$
& $[MM]_{j,x+\hat{l}}$
& $[MMMM]_{jk,x}$\\
$\Delta S^{\tau s}$ &(\protect{\ref{Eq:SeffTS}})
& $\beta_{\tau s}/8d(d-1)$
& $-[V^+V^-]_{j,x}$
& $[MM]_{j,x+\hat{k}}+[MM]_{j,x+\hat{k}+\hat{0}}$\\
\hline
$\Delta \tilde{S}^{\tau}$ &(\protect{\ref{Eq:SeffVV-EHS}})
&$(\bt+\beta_{\tau s}\psi_{\tau s})/4d$
&$-V^+_x$
&$V^-_{x+\hat{j}}$\\
$\Delta \tilde{S}^{s}$ &(\protect{\ref{Eq:SeffMMMM-EHS}})
&$(\beta_s+2\beta_{ss}\bar{\psi}_{ss})/2d(d-1)$
&$[MM]_{j,x}$
&$[MM]_{j,x+\hat{k}}$\\
\hline
\end{tabular}
\end{center}
\end{table}

We first apply the EHS transformation 
in the complex representation,
Eq.~(\ref{Eq:EHSc}),
to NNLO terms in $S_\mathrm{NNLO}$,
Eqs.~(\ref{Eq:SeffTT}), (\ref{Eq:SeffSS}) and (\ref{Eq:SeffTS}).
By substituting $(\alpha, A, B)$ in Eq.~(\ref{Eq:EHSc})
as shown in Table \ref{Table:EHSsubst},
we obtain 
\begin{align}
\Delta S^{\tau\tau}\approx&
\frac {\beta_{\tau\tau}}{2d} \sum_{x,j>0} 
     \left[\psibarTT\psiTT
	-W_x^+\psiTT-\psibarTT W_{x+ \hat{j}}^-
     \right]
	+ (x \leftrightarrow x+\hat{j})
\nn\\
\simeq&
N_\tau L^d \beta_{\tau\tau} \psibarTT\psiTT
-\beta_{\tau\tau}\sum_x(\psiTT W_x^+ + \psibarTT W_x^-)
\label{Eq:SeffTT-EHS}
\ ,\\
\Delta S^{ss}\approx&
\frac {\beta_{ss}}{4d(d-1)(d-2)}
     \sum_{\substack{x,\,j>0,\,|k|>0,\,|l|>0\\|k|\neq j,\,|l|\neq j,\,|l|\neq|k|}}
     \left[
	\bar{\psi}_{ss}\psi_{ss}
	- [MM]_{j,x+\hat{l}}\,\psi_{ss}
	- \bar{\psi}_{ss}[MMMM]_{jk,x}
	\right]
\nn\\
\simeq&
N_\tau L^d \beta_{ss}\bar{\psi}_{ss}\psi_{ss}
-\frac{\beta_{ss}}{d}\psi_{ss}\sum_{x,j>0}[MM]_{j,x}
-\frac{\beta_{ss}\bar{\psi}_{ss}}{d(d-1)}\sum_{x,j>0,k>0,k\not=j}[MMMM]_{jk,x}
\label{Eq:SeffSS-EHS}\\
\Delta S^{\tau s}\approx&
\frac {\beta_{\tau s}}{8d(d-1)} \sum_{x, j > 0, \mid k \mid \neq j}
   \left[
	\bar{\psi}_{\tau s}\psi_{\tau s}
	+ [V^+V^-]_{j,x}\psi_{\tau s}
   \right.
\nn\\
&\qquad\qquad\qquad
   \left.
	- \bar{\psi}_{\tau s}
	([MM]_{j,x+\hat{k}}+[MM]_{j,x+\hat{k}+\hat{0}})
   \right]
	+ (x \leftrightarrow x+\hat{j})
\nn\\
\simeq&
  N_\tau L^d
  \frac{\beta_{\tau s}}{2}
  \bar{\psi}_{\tau s}\psi_{\tau s}
+ \frac{\beta_{\tau s}}{4d}
  \psi_{\tau s}\sum_{x,j>0}[V^+V^-+V^-V^+]_{j,x}
- \frac{\beta_{\tau s}}{d}
  \bar{\psi}_{\tau s}\sum_{x,j>0}[MM]_{j,x}
\ ,
\label{Eq:SeffTS-EHS}
\end{align}
We have introduced the auxiliary fields ($\bar{\psi}_K, \psi_K$)
in $\Delta S^K$ ($K = \tau\tau, \tau s, ss$).
In Eq.~(\ref{Eq:SeffTT-EHS})-(\ref{Eq:SeffTS-EHS}),
the symbol ``$\simeq$'' represents
that constant and isotropic values are assumed
for auxiliary fields ($\bar{\psi}_K, \psi_K$),
and we use this notation in later discussions.
Note that the EHS transformation is applied 
so that the auxiliary fields 
$(\bar{\psi}_{ss},\psi_{ss}, \bar{\psi}_{\tau s}, \psi_{\tau s})$
have no chiral charge
in Eqs.~(\ref{Eq:SeffSS-EHS}) and (\ref{Eq:SeffTS-EHS}).
For example, in Eq.~(\ref{Eq:SeffSS-EHS}),
we have decomposed the six-meson term ($MMMMMM$) 
into $MM$ and $MMMM$ mesonic terms,
and introduced auxiliary fields for their composites,
$(\bar{\psi}_{ss},\psi_{ss})=(\langle MM\rangle,\langle MMMM\rangle)$.

NNLO contributions 
in Eqs.~(\ref{Eq:SeffSS-EHS}) and (\ref{Eq:SeffTS-EHS})
generate terms, $[V^+V^-+V^-V^+]$ and $[MMMM]$.
They can be absorbed into
the NLO contributions, 
Eqs.~(\ref{Eq:SeffT}) and (\ref{Eq:SeffS}).
We put $[V^+V^-+V^-V^+]$ and $[MMMM]$ terms in NLO and NNLO
together and bosonize as
\begin{align}
\Delta \tilde{S}^{\tau}=
&\frac{\bt+\beta_{\tau s}\psi_{\tau s}}{4d}
  \sum_{x,j>0} [V^+V^- + V^-V^+]_{j,x}
\nonumber
\\
\approx&\frac{\beta'_\tau}{4d}
  \sum_{x,j>0}
  \left[
    \bar{\psi}_\tau\psi_\tau+V^+_x\psi_\tau-\bar{\psi}_\tau V^-_{x+\hat{j}}
  \right]
  +(x \leftrightarrow x+\hat{j})
\nn\\
\simeq&
  N_\tau L^d
  \frac{\beta'_\tau}{2}
  \bar{\psi}_\tau\psi_\tau
  +\frac{\beta'_\tau}{2}
   \sum_x (\psi_\tau V^+_x - \bar{\psi}_\tau V^-_x)
\label{Eq:SeffVV-EHS}
\ ,\\
\Delta \tilde{S}^{s}=
&-\frac{\beta_s+2\beta_{ss}\bar{\psi}_{ss}}{2d(d-1)}
\sum_{x,j>0,k>0,k\not=j} [MMMM]_{jk,x}
\nonumber\\
\approx&\frac{\beta'_s}{2d(d-1)}
  \sum_{x,j>0,k>0,k\not=j}
  \left[
   \psibarS\psiS - [MM]_{j,x}\psiS - \psibarS [MM]_{j,x+\hat{k}}
  \right]
\nn\\
\simeq&N_\tau L^d \frac{\beta'_s}{2} \psibarS\psiS
-\frac{\beta'_s}{2d}\sum_{x,j>0}(\psiS+\psibarS)[MM]_{j,x}
\ ,
\label{Eq:SeffMMMM-EHS}
\end{align}
where $\beta'_\tau=\bt+\beta_{\tau s}\psi_{\tau s}$
and $\beta'_s=\beta_s+2\beta_{ss}\bar{\psi}_{ss}$.
We find that
$\Vp\Vm$ and $MMMM$ terms shifts the coefficients 
of SCL effective action terms via EHS transformations.
In previous NLO investigations~\cite{PDevol,Miura:2008gd},
we have developed the method to
express $\Vp\Vm$ and $MMMM$ effects
as modifications of
the wave function renormalization factor, 
quark mass and chemical potential.
In the present investigation,
we need to manipulate
the next-to-nearest neighboring interaction
come from $W^{\pm}$ effects in Eq.~(\ref{Eq:SeffTT-EHS})
before utilizing those techniques developed in NLO.
The $W^{\pm}$ effects are evaluated in the next subsection.

We introduce the chiral condensate auxiliary field $\sigma$
by using the standard HS transformation.
We combine the mesonic hopping $[MM]$ terms
in $S_\mathrm{SCL}$ (Eq.~(\ref{Eq:SeffSCL2})), 
NNLO contributions (Eqs.~(\ref{Eq:SeffSS-EHS}) and (\ref{Eq:SeffTS-EHS})),
and NLO contributions with NNLO effects (Eq.~(\ref{Eq:SeffMMMM-EHS})).
Then we obtain the modified mesonic hopping terms,
which can be bosonized as
\begin{align}
-\frac{\bsigp}{2} \sum_{x,y} M_x V_{xy} M_y
\approx&\bsigp\sum_{x,y}
\left[
 \frac12 \sigma_x V_{xy} \sigma_y + \sigma_x V_{xy} M_y
\right]\nonumber\\
\simeq&N_\tau L^d\,
\frac{\bsigp}{2}\sigma^2 + \bsigp\sigma \sum_x M_x
\label{Eq:SeffSigma}
\ ,\\
\bsigp=&\bsig +
2 \left[ \beta_{ss}\psi_{ss}
+\beta_{\tau s}\bar{\psi}_{\tau s}
+(\beta_s+2\beta_{ss}\bar{\psi}_{ss})(\psiS+\psibarS)
\right]
\ , \\
V_{xy}=&\frac{1}{2d}\sum_j (\delta_{x+\hat{j},y} + \delta_{x-\hat{j},y})
\ .
\end{align}
We have obtained the additional mass term
$\bsigp\sigma\sum_x M_x$.
Combined with the current quark mass $m_0$,
we obtain a constituent quark mass as
\begin{align}
m_q=\bsigp\sigma+m_0\ .\label{Eq:m}
\end{align}

Now we have the NNLO effective action in the spatially local
and bilinear form in the fermion fields.
We collect all the remaining terms
from $S_\mathrm{SCL}$ (Eq.~(\ref{Eq:SeffSCL2})), 
NNLO terms (Eqs.~(\ref{Eq:SeffTT-EHS}), (\ref{Eq:SeffSS-EHS})
and (\ref{Eq:SeffTS-EHS})),
NLO terms (Eqs.~(\ref{Eq:SeffVV-EHS}) and  (\ref{Eq:SeffMMMM-EHS})),
and the chiral condensate term (Eq.~(\ref{Eq:SeffSigma}))
\begin{align}
S_\mathrm{eff}=& \Seff{eff}{F}+\Seff{eff}{X}
\ ,\\
\Seff{eff}{F}=&
\frac12\sum_x \left[
CV^+_x
-\bar{C}V^-_x
\right]
+\sum_x  m_q M_x
-\beta_{\tau\tau} \sum_x (\psiTT W^+_x+\psibarTT W^-_x)
\label{Eq:SeffF}
\ ,\\
C=&1+(\bt+\beta_{\tau s}\psi_{\tau s})\psi_\tau			\ ,\ 
\bar{C}=1+(\bt+\beta_{\tau s}\psi_{\tau s})\bar{\psi}_\tau	\ ,\\
\Seff{eff}{X}=&
N_\tau L^d\left[
\beta_{\tau\tau}\psibarTT\psiTT
+\beta_{ss}\bar{\psi}_{ss}\psi_{ss}
+\frac12\beta_{\tau s}\bar{\psi}_{\tau s}\psi_{\tau s}
+\frac12(\bt+\beta_{\tau s}\psi_{\tau s})\bar{\psi}_\tau\psi_\tau
\right.
\nn\\
&\left.
+\frac12(\beta_s+2\beta_{ss}\bar{\psi}_{ss})\psibarS\psiS
+\frac12 \bsigp\,\sigma^2
\right]
\ ,
\label{Eq:SeffX}
\end{align}
where $S_\mathrm{eff}^{(F)}$ and $S_\mathrm{eff}^{(X)}$ represent
those effective action terms
with and without fermions, respectively.
The SCL and NLO effective action contains $V_x^+, V_x^-$ and $M_x$,
and their corresponding effective potentials are known.\cite{PDevol}
In order to obtain the effective potential 
from the NNLO effective action containing $W_x^+$ and $W_x^-$
in addition to $V_x^+, V_x^-$ and $M_x$,
we discuss how to treat $W_x^+, W_x^-$ terms in the next subsection.

\subsection{Gluonic dressed fermion}
\label{subsec:dressed_fermion}

The fermionic NNLO effective action $S_\mathrm{eff}^{(F)}$
in Eq.~(\ref{Eq:SeffF})
is in the spatially local and bilinear form of $\chi$ and $\chibar$,
then it is possible to obtain the fermion determinant analytically
in the form of Matsubara product.
In order to evaluate the Matsubara product,
we need to obtain the Matsubara frequency which gives zero determinant.
Because of the coupling of 
the next-to-nearest neighboring (NNN) temporal sites
via $W^+$ and $W^-$,
the fermion matrix becomes pentadiagonal rather than tridiagonal
and it is not easy to obtain the solution.
In the present case, NNN terms are in the $\Od(1/g^4)$,
then we can absorb them by introducing a gluonic dressed fermion
\begin{align}
\chi_x' = \chi_x + A e^{\mu} U_{0,x}\,\chi_{x+\hat{0}}
\ ,\quad
\chibar_x' = \chibar_x +\bar{A} \chibar_{x+\hat{0}} e^{-\mu} U^\dagger_{0,x}
\ ,
\end{align}
where $A = \Od(1/g^4)$.
Mesonic composites are represented in this dressed fermion as
\begin{align}
M_x =& M'_x - A V'^+_x - \bar{A} V'^-_x + \Od(1/g^8)\ ,\\
V^+_x =& V'^+_x - \bar{A} M'_{x+\hat{0}} - A W'^+_x
+ \Od(1/g^8)\ ,\\
V^-_x =& V'^-_x - A M'_{x+\hat{0}} - \bar{A} W'^-_x
+ \Od(1/g^8)\ ,\\
W^+_x =& W'^+_x + \Od(1/g^8)\ ,\quad
W^-_x = W'^-_x + \Od(1/g^8)\ ,\quad
\end{align}
where primed composites denote those with dressed fermions,
such as $M'_x = \chibar'_x\chi'_x$.
In terms of this dressed fermion, 
the fermionic term 
$S_\mathrm{eff}^{(F)}$ is rewritten as
\begin{align}
S_\mathrm{eff}^{(F)}
=&\frac12\sum_x\left[(C-2m_qA) V'^+_x - (\bar{C}+2m_q\bar{A}) V'^-_x\right]
+ \sum_x\left[m_q-\frac12(C\bar{A}-\bar{C}A)\right] M'_x
\nn\\
-&\frac12\sum_x \left[
(2\btt\psiTT+CA) W'^+_x+(2\btt\psibarTT-\bar{C}\bar{A})W'^-_x\right]
+\Od(1/g^8)
\ .
\end{align}
We find that NNN terms in the primed fields cancel with each other
when we choose 
$A=-2\beta_{\tau\tau}\psiTT/C$
and
$\bar{A}=2\beta_{\tau\tau}\psibarTT/\bar{C}$.
Then the NNLO fermionic effective action is 
represented in $V'^+_x, V'^-_x$ and $M'_x$ terms.
In the later discussion, we regard $\chi'$ and $\chibar'$
as quarks and anti-quarks,
and omit the prime for the mesonic composites.
The fermionic effective action now reads
\begin{align}
S_\mathrm{eff}^{(F)}=& \frac12\sum_x\left[ \Zm V^+_x - \Zp V^-_x \right]
 + \sum_x m_q^{\prime} M_x 
+\Od(1/g^6)
\label{Eq:SeffDress}
\ ,\\
\Zm= &C-2m_qA = C+4m_q^{\prime}\beta_{\tau\tau}\psiTT +\Od(1/g^6) \ ,\nn\\
\Zp=&\bar{C}+2m_q\bar{A} = \bar{C}+4m_q^{\prime}\btt\psibarTT +\Od(1/g^6) \ ,\\
m_q^{\prime}=&m_q-\frac12(C\bar{A}-\bar{C}A) = m_q-\btt(\psiTT+\psibarTT)+\Od(1/g^6)\ .
\end{align}
Comparing Eq.~(\ref{Eq:SeffDress}) with (\ref{Eq:SeffF}),
we find that the effects of $W^{\pm}$ are expressed as
modifications of the constituent quark mass
and the coefficient of $V^{\pm}$
through the gluonic dressed fermion.
It should be noted that
the Jacobian between $(\chi, \chibar)$ and $(\chi', \chibar')$ 
deviates from unity by $\Od(1/g^{4N_\tau N_c})$,
and we can ignore its effects in NNLO SC-LQCD.
As a result,
NNLO effective action is represented in $V_x^+, V_x^-$ and $M_x$ terms
as in the case of bosonized NLO effective action~\cite{PDevol,Miura:2008gd}.

\subsection{Effective potential}
\label{subsec:Feff}

In the discussion by the previous subsection,
we find that the NLO and NNLO corrections lead to the coefficient modification
of $V^+$, $V^-$ and $M$ in the fermionic effective action.
The meaning of this modification would be more clearly understood
in the following representation of the fermionic effective action
\begin{align}
S_\mathrm{eff}^{(F)}
=& \frac12\sum_x\left[ \Zm V^+_x - \Zp V^-_x \right]
 + \sum_x m_q^{\prime} M_x 
\nn\\
=& Z_\chi \left[
	\sum_{x,y}
	  \frac12\left[
	  	 e^{-\delta\mu}V^+_x
		-e^{\delta\mu}V^-_x
	  \right]
	+\sum_x
	  \mq M_x
	\right]
\nn\\
=& Z_\chi \sum_{xy} \chibar_{x}\,
	G^{-1}_{xy}(\mq;\tilde{\mu},T)\chi_{y}
\ ,\\
G^{-1}_{xy}(\mq;\mu,T)
=& \displaystyle \frac {1}{2} \left[
	 e^{\mu}U_{0,x}\,\delta_{x+\hat{0},y}
	-e^{-\mu}U^\dagger_{0,x}\,\delta_{x-\hat{0},y}
	\right]
   + \mq \delta_{xy}
\ ,\\
Z_\chi = \sqrt{\Zp \Zm}
\ ,&\quad
\mq= \frac{m_q^{\prime}}{Z_\chi}
\ ,\quad
\tilde{\mu}=\mu-\delta \mu=\mu-\log\sqrt{\Zp / \Zm}
\label{Eq:modification} 
\ .
\end{align}
In this form of the effective action,
we easily find that NNLO effects result in the modification of  
the wave function renormalization factor $Z_\chi$,
quark mass $\mq$ and chemical potential $\tilmu$
by using the technique developed in NLO investigations~\cite{PDevol,Miura:2008gd}.

We carry out the Grassmann integral over the quark fields
and the temporal link integral in a standard way~\cite{Damgaard,Faldt:1985ec}.
First, we perform the Fourier transformation in the temporal coordinate,
and obtain the product in the frequency by the Grassmann integral
over the quark fields.
Second, we evaluate the product in the frequency by using the Matsubara method.
Finally, we carry out the temporal link integral by using the Haar measure.
The fermionic part in the effective potential 
$\Fq$ is given as
\begin{align}
&\Fq
\equiv -\frac{1}{N_\tau L^d}\log\left[\int {\cal D}[\chi,\chibar,U_0]
	e^{-\Seff{eff}{F}}
	\right]
={\cal V}_q(\mq;\tilde{\mu},T) - N_c\log Z_\chi
\ ,\\
&\Vq(\mq;\tilde{\mu},T)
=-T\log\left[
	    \frac {\sinh[(N_c+1)E_q(\mq)/T]}{\sinh[E_q(\mq)/T]}
		+2\cosh(N_c\tilmu/T)
	\right]
\label{Eq:FeffMq}
\ , 
\end{align}
where the number of temporal sites is replaced as $N_\tau=1/T$,
and $E_q(\mq)=\mathrm{arcsinh}\,(\mq)$
is regarded as the quark excitation energy.

The latter point is understood from the expression of
the temporal quark hopping matrix determinant,\cite{Faldt:1985ec}
\begin{align}
e^{-{\cal V}_qL^d/T}
&=\int {\cal D}[U_0]\, \mathrm{det}_{x,c}\,
	G_{xy}^{-1}(\tilde{m}_q;\tilde{\mu}, T)
\label{Eq:explainEq0} \\
&=\prod_{\mathbf{x}}\int d\mathcal{U}_0\, \mathrm{det}_c \left[
	 e^{E_q/T}
	\left(1+\mathcal{U}_0 e^{-(E_q-\tilde{\mu})/T}\right)
	\left(1+\mathcal{U}_0^{\dagger} e^{-(E_q+\tilde{\mu})/T}\right)
	\right]
\label{Eq:explainEq}
\ ,
\end{align}
where $\mathcal{U}_0(\mathbf{x})=\prod_{\tau}U_0(\mathbf{x},\tau)$.
We can evaluate this temporal link integral exactly,
for example, in the Polyakov gauge,\cite{Damgaard}
\begin{align}
U_{0}(\mathbf{x},\tau)
=
\mathrm{diag}
(e^{iT\theta_1(\mathbf{x})},\cdots,e^{iT\theta_{N_c}(\mathbf{x})})
\ .\label{Eq:Tgauge}
\end{align}
We find the Boltzmann factor $\exp\left[-(E_q\pm\tilde{\mu})/T\right]$
appears, and the integrand is nothing but the thermal quark contribution
to the partition function if $U_0$ is close to unity.
Note that the determinant
is for the spacetime and color in Eq.~(\ref{Eq:explainEq0}), 
and for the color in Eq.~(\ref{Eq:explainEq}).

We obtain the NNLO effective potential as follows.
\begin{align}
\Feff
\equiv& -\frac{1}{N_\tau L^d}\log\left[\int {\cal D}[\chi,\chibar,U_0]
	e^{-\Seff{eff}{F}-\Seff{eff}{X}}
	\right]
\nn\\
=&\Fq(\Phi;\mu,T)
+\Feff^{(X)}(\Phi)
\label{Eq:Feff}
\ ,\\
\Feff^{(X)}
=& \Seff{eff}{X}/N_\tau L^d
\nn\\
=& 
 \frac12 \bsigp \,\sigma^2
+\frac12(\bt+\bts\psi_{\tau s})\,\psibarT\psiT
+\frac12(\bs+2\bss\psibarSS)\bar{\psi}_s\psi_s
\nn\\
&+\btt\psibarTT\psiTT
+\bss\psibarSS\psiSS
+\frac12\bts\psibarTS\psiTS
\ ,\\
\mq=& \frac{m_q^{\prime}}{Z_\chi}
\ ,\quad
m_q^{\prime}= \bsigp \sigma + m_0 - \btt(\psibarTT+\psiTT)
\ ,\\
\Zp=& 1+(\bt+\bts\psiTS)\psibarT+4\btt m_q^{\prime}\psibarTT
\ ,\\
\Zm=& 1+(\bt+\bts\psiTS)\psiT   +4\btt m_q^{\prime}\psiTT
\ ,\\
\bsigp=& \bsig
	+(\bs+2\bss\bar{\psi}_{ss})(\psi_s+\bar{\psi}_s)
	+2\bss\psi_{ss}+2\bts\bar{\psi}_{\tau s}
\ ,
\end{align}
where $\Phi=(\sigma,\{\psi_K, \bar{\psi}_K; K=\tau,s,\tau\tau,\tau s,ss \})$.
Coupling constants
($b_\sigma, \bt, \beta_s, \beta_{\tau\tau}, \beta_{\tau s}, \beta_{ss}$)
are defined in Eqs.~(\ref{Eq:CoeffA}) and ~(\ref{Eq:CoeffB}).

\subsection{Stationary conditions}

The auxiliary fields introduced during the bosonization procedure
have to satisfy the stationary condition,
$\partial\Feff/\partial\Phi=0$.
Within the constant field approximation,
the stationary values of auxiliary fields are related with each other,
then we replace the auxiliary fields other than 
($\sigma$, $\psibarT$ and $\psiT$)
with their equilibrium values as functions of
($\sigma$, $\psibarT$ and $\psiT$).

We can solve the stationary conditions for all auxiliary fields exactly.
In solving the stationary condition,
we first note that the quark free energy is a function of
$m_q^{\prime}$, $\Zp$ and $\Zm$, as we can guess from 
Eq.~(\ref{Eq:SeffDress}). 
Then the variation $\Fq$ is obtained as
\begin{align} 
\frac{\partial\Fq}{\partial \Phi}
=\frac{\partial\Fq}{\partial m_q^{\prime}} \frac{\partial m_q^{\prime}}{\partial\Phi}
+ \frac{\partial\Fq}{\partial \Zp} \frac{\partial \Zp}{\partial\Phi}
+ \frac{\partial\Fq}{\partial \Zm} \frac{\partial \Zm}{\partial\Phi}
\label{Eq:Fq_derivative}
\ . 
\end{align} 
Since $\Zp$ and $\Zm$ contain $m_q^{\prime}$,
the above derivative $\partial\Fq/\partial m_q^{\prime}$ also contains
the derivative via $\Zpm$,
$(\partial\Fq/\partial\Zpm)(\partial\Zpm/\partial m_q^{\prime})$.
Substituting $\sigma, \psiS, \psibarS, \psibarT, \psiT$, 
in Eq.~(\ref{Eq:Fq_derivative}), 
the stationary conditions for these fields read,
\begin{align} 
\label{Eq:CondA}
 \sigma=-\frac{\partial\Fq}{\partial m_q^{\prime}}	\ ,\quad
 \psi_s=\bar{\psi}_s=\sigma^2				\ ,\quad
 \psibarT=-2 \frac{\partial \Fq}{\partial\Zm}	\ ,\quad
 \psiT   =-2 \frac{\partial \Fq}{\partial\Zp}	\ .
\end{align} 
By using these relations,
the stationary conditions for NNLO auxiliary fields
($\psibarTT$, $\psiTT$, $\psibarTS$, $\psiTS$, $\psibarSS$, $\psiSS$)
are obtained as
\begin{align}
&\psibarTT =2m_q^{\prime} \psibarT - \sigma \ ,\quad
 \psiTT=2m_q^{\prime} \psiT - \sigma
\label{Eq:stat-NNLO-A}	\ ,\\
&\psibarTS=\psibarT\psiT	\ ,\quad
 \psiTS=2\sigma^2		\ ,\quad
 \psibarSS=\sigma^2		\ ,\quad
 \psiSS=\sigma^4		\ .
\label{Eq:stat-NNLO-B}
\end{align} 
From Eq.~(\ref{Eq:stat-NNLO-A}),
we obtain $m_q^{\prime}$ as a function of $(\sigma,\psiT,\psibarT)$
\begin{align}
m_q^{\prime} = \frac{\bsigp\,\sigma +m_0+2\btt\sigma}{1+2\btt(\psiT+\psibarT)}\ .
\label{Eq:mprime}
\end{align}
Now all the stationary values of auxiliary fields
other than $(\sigma,\psiT,\psibarT)$
are obtained as functions of these three fields.

We shall now proceed to erase one more auxiliary field.
Three derivatives in Eq.~(\ref{Eq:Fq_derivative}),
$\partial\Fq/\partial m_q^{\prime}$ and $\partial\Fq/\partial\Zpm$, 
are not independent, but are related to the two derivatives,
$\partial\Vq/\partial\mq$ and $\partial\Vq/\partial\tilmu$.
The stationary conditions for $\psibarT$ and $\psiT$
in Eq.~(\ref{Eq:CondA})
are rewritten in terms of these derivatives as
\begin{align}
\Zm \psibarT
    =& \frac{m_q^{\prime}}{\Zchi} \frac{\partial \Vq}{\partial \mq}
    -\frac{\partial \Vq}{\partial \tilmu} + N_c
\label{Eq:Cond_psibarTT}
\, \\
\Zp \psiT
    =& \frac{m_q^{\prime}}{\Zchi} \frac{\partial \Vq}{\partial \mq}
    +\frac{\partial \Vq}{\partial \tilmu} + N_c
\label{Eq:Cond_psiTT}
\ . 
\end{align}
We solve the coupled equations,
Eqs.~(\ref{Eq:Cond_psibarTT}) and (\ref{Eq:Cond_psiTT}), 
for $\partial\Vq/\partial\mq$ and $\partial\Vq/\partial\tilmu$
\begin{align}
\frac{\partial\Vq}{\partial\tilde{\mu}}=& -\omega_\tau (1-4m_q^{\prime}\btt\sigma)
\label{Eq:Cond_wt}
\ , \\
\frac{m_q^{\prime}}{Z_\chi}\,\frac{\partial\Vq}{\partial\mq}
=&\frac12(\Zp\psiT+\Zm\psibarT)-N_c
\nn\\
=&\phiT+\bt'\psibarT\psiT+2\btt m_q^{\prime}(\psibarTT\psiT+\psiTT\psibarT)
	-N_c
\label{Eq:Cond_phiT}
\ . 
\end{align}
We have parameterized the auxiliary fields $\psibarT,\psiT$ as
$\psibarT=\phiT+\wt$ and $\psiT=\phiT-\wt$.
This corresponds to applying the EHS transformation in Eq.~(\ref{Eq:EHSb}),
where we substitute $\phi \to i\omega$.
Equation~(\ref{Eq:Cond_wt}) implies that $\wt$ is related
to the quark number density,
$\rho_q \equiv -\partial\Feff/\partial\mu = -\partial\Vq/\partial\tilmu$.
It is also possible to solve the stationary condition for $\sigma$
in Eq.~(\ref{Eq:CondA}) for $\partial\Vq/\partial\mq$.
\begin{align}
\frac{1}{Z_\chi}\,\frac{\partial\Vq}{\partial\mq}
=&-\sigma+2\btt(\psibarTT\psiT+\psiTT\psibarT)
\label{Eq:Cond_sigma}
\ ,
\end{align}
Two equations Eqs.~(\ref{Eq:Cond_phiT}) and (\ref{Eq:Cond_sigma})
have to be consistent, then we find the relation between 
$\sigma$, $\wt$ and $\phiT$
\begin{align}
\phiT+\bt'(\phiT^2-\wt^2)+m_q^{\prime}\sigma-N_c=0\ .
\end{align}
By substituting Eq.~(\ref{Eq:mprime}),
we obtain a cubic equation for $\phiT$
\begin{align}
4\bt'\btt\phiT^3
+(\bt'+4\btt+2\bts \sigma^2)\phiT^2
+(1-4N_c\btt-4\bt'\btt\wt^2)\phiT&
\nn\\
-N_c+\sigma[(\bsig + 2\btt + 2\bs\sigma^2 + 6\bss\sigma^4
-2\bts\wt^2)\sigma+m_0]-\bt'\wt^2
&=0
\ ,
\label{Eq:stat-phiT}
\end{align}
and its solution $\phiT=\phiT(\sigma,\wt)$ is obtained analytically.

We have obtained the final expression of
the effective potential as a function of the three auxiliary fields 
$(\sigma,\phiT,\wt)$,
temperature $T$ and chemical potential $\mu$
\begin{align}
\Feff
=&\Feff^{(X)}+{\cal V}_q(\mq;\tilde{\mu},T) -N_c\log \sqrt{\Zp\Zm}
\label{Eq:Feff_final}
\ ,\\
\Feff^{(X)}
=& 
 \frac12 \bsigt \,\sigma^2+\frac12\beta_s\sigma^4+2\bss\sigma^6
\nn\\
&+\frac12(\bt+4\bts\sigma^2+8\btt {m_q^{\prime}}^2)(\phiT^2-\omega_\tau^2)
-4\btt m_q^{\prime}\phiT\sigma
\ ,\\
m_q^{\prime}=&\frac{\bsigt \sigma + m_0}{1+4\btt\phiT}
\ ,\quad
\mq=\frac{m_q^{\prime}}{\sqrt{\Zp\Zm}}
\label{Eq:mq_final}
\ ,\\
Z_{\pm} =& 1+(\bt+2\bts\sigma^2+8\btt {m_q^{\prime}}^2)(\phiT \pm \wt)
	- 4\btt m_q^{\prime}\sigma
\ ,\\
\bsigt =& b_{\sigma}^{\prime}+2\beta_{\tau\tau}
	=\bsig
	+2\btt
	+2\bs\sigma^2+6\bss\sigma^4
	+2\bts(\phiT^2-\wt^2)
\ ,\\
\tilmu=&\mu-\log\sqrt{\Zp/\Zm}
\ . 
\end{align}

Since $\phiT$ is a function of $\sigma$ and $\wt$ whose function form is
independent from $(T,\mu)$, we cannot regard it as an order parameter.
Thus, we have two order parameters in the present treatment of NNLO SC-LQCD,
the chiral condensate $\sigma$ and $\wt$.
We can regard $\wt$ is a vector potential field for quarks;
the chemical potential shift is mainly determined by $\wt$,
and $\wt$ contributes repulsively to the effective potential in equilibrium.
This two order parameter feature may be a natural consequence
from the potential term from quarks, $\Vq(\mq;\tilmu,T)$. 
There are two independent derivatives,
$\partial\Vq/\partial\mq$ and $\partial\Vq/\partial\tilmu$,
which appear in the equilibrium condition,
then we have two degrees of freedom.

\section{Chiral Phase Transition in NNLO SC-LQCD}
\label{Sec:Results}

The effective potential derived in the previous section
determines the equilibrium (vacuum) and the phase structure of QCD matter.
In this section, we investigate the chiral phase transition
in the chiral limit $m_0=0$ at $N_c=3$.
First, we discuss the effective potential surface and the stationary conditions.
Then, critical temperature and chemical potential are investigated.
Finally, we discuss the coupling dependence of the critical point.

\subsection{Effective potential surface}
\label{subsec:Feff_surface}

\begin{figure}[htb]
\begin{center}
\Psfig{6cm}{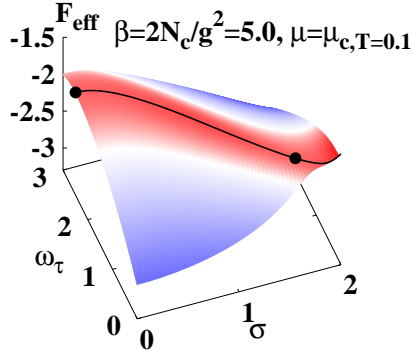}
\caption{The effective potential as a function of $\sigma$ and $\wt$ 
at ($T,\mu,\beta$)=$(0.1,\mu_c,5.0)$ in the lattice unit.
The stationary condition 
for $\wt$ obtained in Eq.~(\protect\ref{Eq:Cond_wt})
is satisfied on the solid curve.
}
\label{Fig:Feff_surface}
\end{center}
\end{figure}

\begin{figure}[htb]
\begin{center}
\Psfig{\hsize}{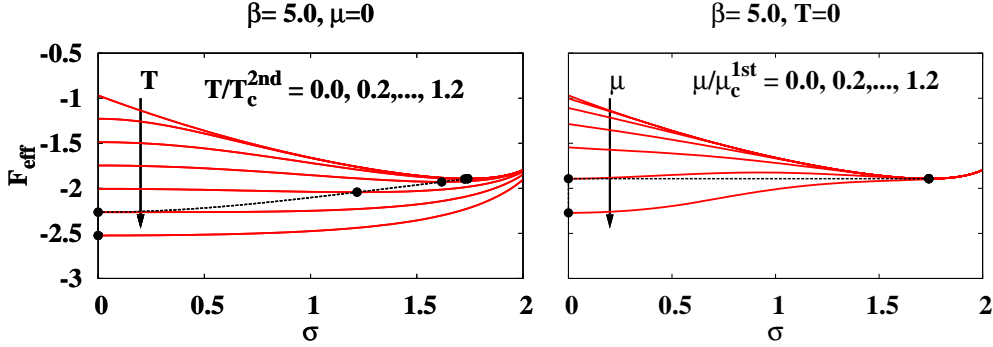}
\caption{The effective potential
in the lattice unit is illustrated
as a function of the chiral condensate $\sigma$
on the $T$-axis (left) and the $\mu$-axis (right) 
for $\beta=5.0$.
The filled circles show the equilibrium points.
In the left (right) panel, the effective potential decreases
for increasing $T(\mu)$.
}
\label{Fig:Feff_Tdep-and-Feff_mudep}
\end{center}
\end{figure}

The equilibrium is determined 
by the stationary condition of $\Feff$ with respect to the auxiliary fields.
By substituting $\phiT$ in Eq.~(\ref{Eq:Feff_final})
with the solution of Eq.~(\ref{Eq:stat-phiT}),
we obtain the effective potential as a function of $\sigma$ and $\wt$,
$\Feff(\sigma,\wt)=\Feff(\sigma,\wt,\phiT=\phiT(\sigma,\wt))$.
The remaining stationary conditions are for $\sigma$ and $\wt$
\begin{align}
\displaystyle \frac {\partial \Feff}{\partial \sigma}
=& \displaystyle \frac {\partial \Feff}{\partial \wt} = 0
\ ,
\end{align}
and we have to solve this coupled equation self-consistently.
This is equivalent to searching 
for the saddle point of $\Feff$~\cite{Bilic,PDevol,Miura:2008gd},
where $\Feff$ is convex downward and upward in $\sigma$ and $\wt$ directions,
$\partial^2\Feff/\partial\sigma^2>0$ and $\partial^2\Feff/\partial\wt^2<0$.
In Fig.~\ref{Fig:Feff_surface}, 
we display $\Feff(\sigma,\wt)$ at $(T,\mu,\beta)=(0.1, \mu_{c,T=0.1}, 5.0)$,
as an example.
The solid curve in Fig.~\ref{Fig:Feff_surface}
shows the stationary condition for $\wt$, $\wt=\wt^\mathrm{stat.}(\sigma)$,
and the filled circles show the equilibrium points.
The effective potential as a function of $\sigma$
for given $(T,\mu)$ is defined
as $\Feff(\sigma)=\Feff(\sigma,\wt=\wt^\mathrm{stat.}(\sigma))$,
whose minimum corresponds to the equilibrium.
In the left (right) panel of Fig.~\ref{Fig:Feff_Tdep-and-Feff_mudep},
we display $\Feff(\sigma)$ on the $T$-($\mu$-)axis.
The filled circles show the equilibrium points,
which are obtained from the stationary condition for $\sigma$,
Eq.~(\ref{Eq:CondA}).

In vacuum,
the chiral condensate is non-zero,
{\em i.e.} the chiral symmetry is spontaneously broken
and the equilibrium is in the NG phase. 
As temperature increases on the $T$-axis,
the chiral condensate decreases smoothly and become zero at $T=T_{c,\mu=0}$,
{\em i.e.} the chiral phase transition to the Wigner phase takes place.
As chemical potential increases on the $\mu$-axis,
the chiral condensate changes little and jump to zero at $\mu=\mu_{c,T=0}$.
The chiral phase transitions are the second- and the first-order
on the $T$- and the $\mu$-axes, respectively, at $\beta = 5.0$.
These results are consistent with those
in SCL\cite{Bilic,Nishida:2003uj,Fukushima:2003vi,Nishida:2003fb,deForcrand:2009dh} and NLO~\cite{PDevol} SC-LQCD.

\subsection{Critical temperature and chemical potential}
\label{subsec:Tcmuc}

The second-order phase transition boundary is obtained from the condition
of $C_2=0$ where $\Feff(\sigma)=\sum_n C_n \sigma^n / n!$.
In order to obtain $C_2$,
we start from the effective potential as a function of $\sigma,\phiT$ and $\wt$,
$\Feff(\sigma,\wt,\phiT)$ in Eq.~(\ref{Eq:Feff_final}).
From the stationary conditions,
$\phiT$ and $\wt$ are found to be even functions of $\sigma$
in the chiral limit 
($\left. \partial \Phi'/\partial \sigma \right|_{\sigma=0}=0$,
 $\Phi'=\phiT, \wt$).
Because of the stationary condition, 
the first derivative of $\Feff$ by auxiliary fields are zero
($\partial \Feff/\partial \Phi=0$).
By using these properties, $C_2$ is expressed as, 
\begin{align}
C_2
 =& \left. 
\left( \displaystyle \frac {\partial}{\partial \sigma} 
+ \sum_{\Phi'=\phiT,\wt} 
\displaystyle \frac {\partial \Phi'}{\partial \sigma} 
\displaystyle \frac {\partial}{\partial \Phi'} 
\right)^2 \Feff
\right|_{\sigma=0}
 = \left.
\displaystyle \frac {\partial^2 \Feff}{\partial \sigma^2}
\right|_{\sigma=0}
\label{Eq:C2a}
\\
=& C_2^{(X)} + C_2^{(\Vq)} - \displaystyle \frac {N_c}{2}
\left[ \displaystyle \frac {C_2^{(\Zp)}}{\Zp}
+ \displaystyle \frac {C_2^{(\Zm)}}{\Zm} \right]
\ ,
\label{Eq:C2}
\end{align}
where, 
\begin{align}
C_2^{(X)}
=& \frac{\partial^2\Feff^{(X)}}{\partial\sigma^2}
= \bsigt + \left( 4 \bts 
+ 8 \btt C_{m_q^{\prime}}^2
\right) (\phiT^2-\wt^2)
- 8 \btt \phiT C_{m_q^{\prime}}
\ ,\\
C_2^{(\Vq)}
=& \frac{\partial^2\Vq}{\partial\sigma^2}
= \frac{\rho_q}{2} 
\left[ \Disp{\frac{C_2^{(\Zp)}}{\Zp}} - \Disp{\frac{C_2^{(\Zm)}}{\Zm}} \right]
- \Disp{\frac{N_c(N_c+1)(N_c+2)}{3T(2\cosh(N_c \tilmu/T) + N_c+1)}}
\cdot \left( \displaystyle \frac {C_{m_q^{\prime}}}{ \Zchi} \right)^2
\ ,\\
C_2^{(\Zpm)}
=& \frac{\partial^2\Zpm}{\partial\sigma^2}
= \left( 4 \bts 
+ 16 \btt C_{m_q^{\prime}}^2
\right) 
(\phiT \pm \wt)
- 8 \btt C_{m_q^{\prime}}
\ , \\
C_{m_q^{\prime}}
=&\frac{m_q^{\prime}}{\sigma}
=\frac{\bsigt}{1 + 4 \btt \phiT}
\ .
\label{Eq:Cm}
\end{align}
In Eqs.(\ref{Eq:C2})--(\ref{Eq:Cm}),
the right hand sides have to be evaluated
in the conditions $\sigma=0$
and $\rho_q = -\partial \Vq / \partial \tilmu$.

\begin{figure}[bth]
\begin{center}
\Psfig{10cm}{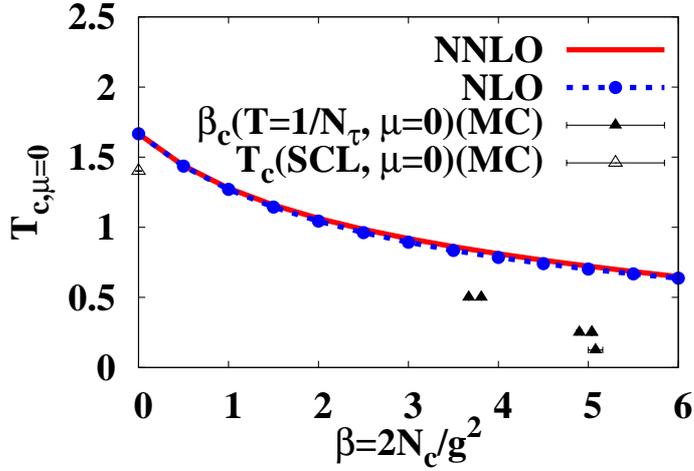} 
\caption{The $\beta$ dependences
of the critical temperature at $\mu=0$ ($T_{c,\mu=0}$)
in the lattice unit.
We compare
the NLO (the dashed line with circles) 
and NNLO (the solid line) results.
The triangles represent
the results of
the critical temperature ($T_{c,\mu=0}$, open triangle)
and the critical coupling ($\beta_c$, filled triangles)
obtained in Monte-Carlo simulations
with one species of unrooted staggered fermion: 
From the left,
$T_{c,\mu=0}$ in the SCL with MDP simulations~\cite{deForcrand:2009dh},
$\beta_c$ at $(N_\tau,m_0)=(2,0.025)$,\cite{Forcrand}
$(2,0.05)$,\cite{Forcrand},
$(4,0.0)$,\cite{Gottlieb:1987eg}
$(4,0.05)$,\cite{D'Elia:2002gd,Fodor:2001au} 
and $(8,0.0)$.\cite{Gavai:1990ny}
}
\label{Fig:Tc-NNLO-A-NLO-A}
\end{center}
\end{figure}

\begin{figure}[bth]
\begin{center}
\Psfig{\hsize}{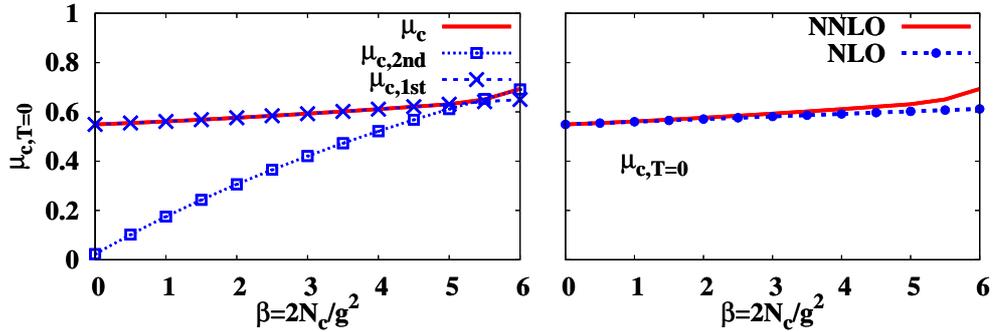} 
\caption{The $\beta$ dependences
of the critical chemical potential at $T=0$ ($\mu_{c,T=0}$)
in the lattice unit.
In the left panel, 
we show NNLO results for
the critical chemical potential
(the solid line),
the first-order critical chemical potential
(the dashed line with crosses) 
and the second-order critical chemical potential
(the dotted line with squares).
In the right panel, we compare
the NLO (the dashed line with circles) 
and NNLO (the solid line) results.
}
\label{Fig:muc-NNLO-A-and-muc-NNLO-A-NLO-A}
\end{center}
\end{figure}

The second order critical temperature at $\mu=0$ is obtained
from the condition $C_2=0$ at $\mu=0$ as
\begin{align}
T_{c,\mu=0}
=T_{c}^{(\mathrm{SCL})} \cdot \displaystyle \frac {C_{m_q^{\prime}}^2}{\bsig \Zchi^2}
\left( C_2^{(X)}-\displaystyle \frac{N_c C_2^{(Z_+)}}{Z_+} \right)^{-1}
\ , 
\end{align}
where $T_{c}^{(\mathrm{SCL})}= d(N_c+1)(N_c+2)/[6(N_c+3)]$
denotes the critical temperature at $\mu=0$ in the strong coupling limit.
Note that at $\mu=0$, $\wt$ becomes zero (i.e. $\tilmu=0$).
Therefore, the coefficients in $T_{c,\mu=0}$ are the value at $\wt=0$,
and then we find $Z_+=Z_-$ ($C_2^{Z_+}=C_2^{Z_-}$).
$T_{c,\mu=0}$ decreases as $\beta$ increases because of finite coupling effect.

In Fig.~\ref{Fig:Tc-NNLO-A-NLO-A},
we compare the critical temperature in NLO and NNLO.
In both NLO and NNLO, $T_{c,\mu=0}$ decreases as $\beta$ increases,
and $T_{c,\mu=0}$ in NLO and NNLO exhibit almost the same values.
%
We also show the MC results
with one species of unrooted staggered fermion
on the critical coupling ($\beta_c$)
for given $N_\tau=1/T=2, 4~\mathrm{and}~8$ at $\mu=0$ (filled triangles)
and the critical temperature in SCL:
From the left,
we show $T_{c,\mu=0}$
in the SCL with monomer-dimer-polymer (MDP)
simulations~\cite{deForcrand:2009dh},
$\beta_c$ at
$N_\tau=2$ (the quark mass $m_0=0.025$),\cite{Forcrand}
$N_\tau=2$ ($m_0=0.05$),\cite{Forcrand}
$8^3\times 4$ lattice (with a chiral extrapolation),\cite{Gottlieb:1987eg}
$8^3\times 4$ lattice ($m_0=0.05$),\cite{D'Elia:2002gd}
$6^3\times 4$ lattice ($m_0=0.05$),\cite{Fodor:2001au}
and 
$N_\tau=8$ (with a chiral extrapolation).\cite{Gavai:1990ny}
The reduction of $T_{c,\mu=0}$ is not enough to explain the MC results.
In this formulation, we do not consider the deconfinement phase transition 
and the quarks are confined.
$T_{c,\mu=0}$ could decrease than NNLO results
from the effects of the Polyakov loop which is an order parameter
of the deconfinement phase transition~\cite{NNLOwithPolyakov}.
In NNLO, the phase transition at $\mu=0$ stays to be the second order
in the region $\beta\leq6$.
This point differs from the numerical simulation results,
where the phase transition at $\mu=0$
is shown to be the first order
in the continuum region.\cite{D'Elia:2002gd}
The relation with the critical point position
will be discussed in the next subsection.

\begin{figure}[bth]
\begin{center}
\Psfig{\hsize}{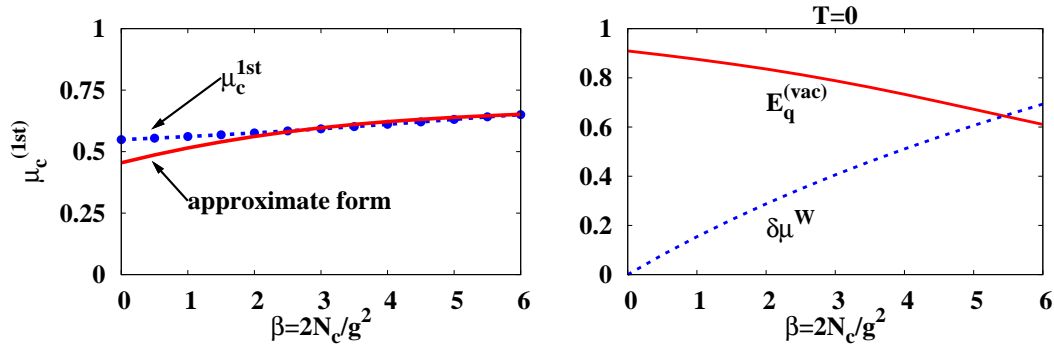} 
\caption{
In the left panel, we show the critical chemical potential 
at $T=0$ ($\mu_c^{(1st)}$,~the dashed line with circles).
The result obtained by ``approximate expression'', 
$(E_q^{(vac)}+\delta\mu^{W})/2$,
is also shown for a comparison (the solid line).
In the right panel, we compare at $T=0$
a mass modification 
$E_q^{(vac)}$
(the upper solid line) with 
a chemical potential modification $\delta\mu^{W}$
(the lower dashed line) as a function of $\beta$.
In both panels, we use the lattice unit.
}\label{Fig:muc-approx-and-competition-bet-dmq-deltamu}
\end{center}
\end{figure}

In Fig.~\ref{Fig:muc-NNLO-A-and-muc-NNLO-A-NLO-A},
the phase transition on $\mu$-axis is numerically found to be the first-order
for $\beta \lesssim 5.5$.
We find that $\mu_{c,T=0}^{(1st)}$ is not largely affected
by finite coupling effects.
This is understood as follows.
Taking the limit $T\to 0$ in Eq.~(\ref{Eq:FeffMq}),
one can easily derive that the quark free energy reduces to
\begin{align}
\Vq = & \left\{
\begin{array}{cc}
-N_c E_q & (E_q \geq \tilmu) \\ 
-N_c \tilmu & (E_q \leq \tilmu) \\ 
\end{array}
\right.
\ .\label{T0Vq}
\end{align}
As explained after Eq.~(\ref{Eq:explainEq}),
$E_q$ represents the quark excitation energy.
In this formulation, we do not consider the deconfinement phase transition and the
quarks are confined. Then, the one and two quark excitation contribution
disappears after temporal link ($U_0$) integration.
As a result, the factor $N_c$ accompanies $E_q$ in Eq.~(\ref{T0Vq}).
This means that the contribution from three co-moving quarks appear as a baryon in the effective 
potential at low $T$.

At $\mu=\mu_{c,T=0}^\mathrm{(1st)}$, effective potentials at two local minima
are the same,
then the following relation holds
\begin{align}
-\frac12 N_c E_q^\mathrm{NG}
& + \left( -\frac12 N_c E_q^\mathrm{NG} +\Feff^{(X),\mathrm{NG}} 
- N_c \log Z_\chi^{\mathrm{NG}} \right)
\nn\\
=-N_c \mu
& + \frac12 N_c \delta\mu^\mathrm{W}
 + \left( \frac12 N_c \delta\mu^\mathrm{W} + \Feff^{(X),\mathrm{W}} 
 - N_c \log Z_\chi^{\mathrm{W}} \right)
\ ,\label{Eq:PotEq0}
\end{align}
where the lhs and rhs show the effective potentials
in the Nambu-Goldstone (NG) and the Wigner phases, respectively.
We numerically find that the lhs and rhs in the brackets 
have almost the same values.
Then, the above relation is approximately represented as
\begin{align}
-\frac12 N_c E_q^\mathrm{NG} \simeq 
-N_c \mu + \frac12 N_c \delta\mu^\mathrm{W}\ ,\label{Eq:PotEq}
\end{align}
These approximate expressions are explained 
based on the two-body interaction
dominance, where the total potential energy amounts to be half
of the single particle potential.
This relation is exact when the effective potential is a quadratic
function of $\sigma$ or $\wt$.
In the NG phase, the quark number density is very small at $T=0$,
then the main contribution to the potential energy comes from $\sigma$.
In the Wigner phase, the chiral symmetry is restored,
then the effective potential becomes a function of $\wt$.
Since there are contributions from other auxiliary fields and 
the quark free energy is not a linear function of $\sigma$ or $\wt$,
the above relation is an approximate one.
In the NG phase at $T=0$, 
the quark number density is almost zero,
and the quark excitation energy 
is almost the same as that in vacuum, 
$E_q^\mathrm{NG}(\mu>0)\simeq E_q^\mathrm{NG}(\mu=0)\equiv E_q^{(\mathrm{vac})}$.
In addition, we obtain 
$\omega_{\tau}=\rho_q \simeq N_c$
in the Wigner phase at $T=0$.
By using these facts,
the Eq.~(\ref{Eq:PotEq}) reduces to,
\begin{gather}
\mu_{c,T=0}^\mathrm{(1st,App.)}
=\frac12\left[
	  E_q^\mathrm{(vac)}
	+ \delta\mu^\mathrm{W}
	\right]
\ ,\label{Eq:muc_approx}\\
\delta\mu^\mathrm{W}=\delta\mu(\sigma=0,\wt=N_c)
\ .\label{Eq:muc_approxB}
\end{gather}
In the left panel of Fig.~\ref{Fig:muc-approx-and-competition-bet-dmq-deltamu},
we compare the first order critical chemical potential
with its approximate expression in Eq.~(\ref{Eq:muc_approx}).
For $\beta \gtrsim 3$,
$\mu_{c,T=0}^\mathrm{(1st,App.)}$ can roughly explain $\mu_{c,T=0}^\mathrm{(1st)}$.

In the right panel of Fig.~\ref{Fig:muc-approx-and-competition-bet-dmq-deltamu},
we show the 
$\beta$ dependence of $E_q$ and $\delta\mu$.
With increasing $\beta$, $E_q$ becomes smaller but 
$\delta\mu^\mathrm{W}$ goes to a larger value.
The sum of them appearing in Eq.~(\ref{Eq:muc_approx}) slightly increases 
and is not very sensitive to $\beta$.
Thus, the quark mass (or excitation energy) and chemical potential
suppressions which come from the finite $\beta$ effects
cancel each other out, and we observe
a small modification of the critical chemical potential.

\subsection{Critical point evolution}
\label{subsec:CEP-evolution}

\begin{figure}[bth]
\begin{center}
\Psfig{10cm}{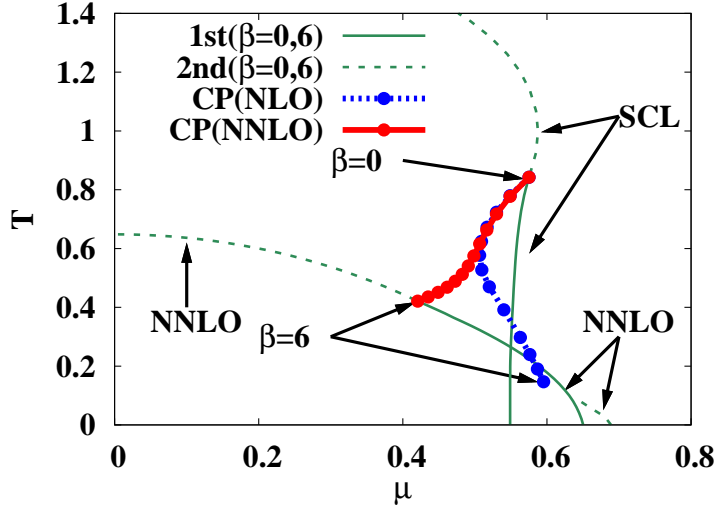} 
\caption{The critical point (CP) flow 
resulting from increasing $\beta$ in $T-\mu$ plane in the lattice unit.
The solid and the dotted lines
represent trajectories of NNLO and NLO cases, respectively.
We also show the first and second-order phase transition boundaries
obtained in cases of SCL and NNLO at $\beta=6$.}
\label{Fig:CEP-evolution}
\end{center}
\end{figure}

\begin{figure}[bth]
\begin{center}
\Psfig{15cm}{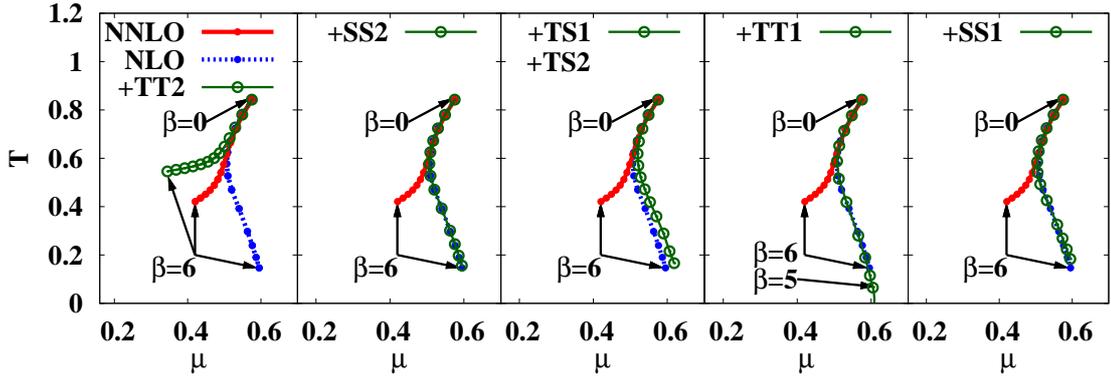} 
\caption{The critical point flow 
with increasing $\beta$ in several truncation schemes in the lattice unit.
From the left panel,
a part of NNLO effects,
NNLO-TT2, NNLO-SS2, NNLO-TS12,
NNLO-TT1 or NNLO-SS1 is taken into account.
We compare them with 
NLO (the dotted line with circles)
and full NNLO (the solid line with circles) results.
}
\label{Fig:CEP-evolution-truncation-NNLO-A}
\end{center}
\end{figure}

\begin{figure}[bth]
\begin{center}
\Psfig{10cm}{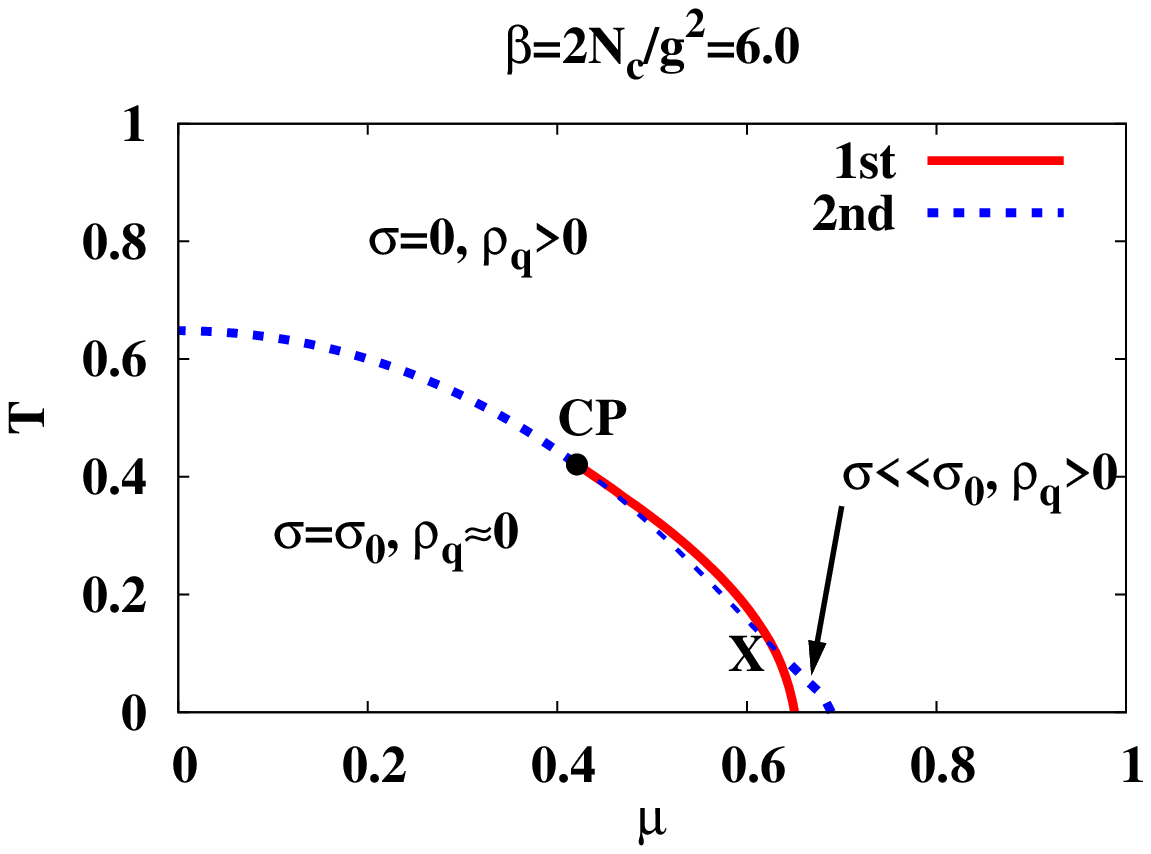} 
\caption{The NNLO phase diagram at $\beta=6.0$ in the lattice unit.
The solid and thick dashed lines 
represent the first- and second-order 
phase transition boundaries, respectively.
The filled circle represents the tri-critical point.
On the (thick and thin) dashed line,
$C_2=0$ is satisfied, where $C_2$ is the curvature of the effective potential
in Eq.~(\protect{\ref{Eq:C2a}}).
Note that there is no phase transition on the thin dashed line
between CP and X.
By contrast, in the region where $T$ is lower than that of X,
two sequential phase transitions occur as $\mu$ becomes large.}
\label{Fig:phase-diagram}
\end{center}
\end{figure}

In Fig.~\ref{Fig:CEP-evolution}, 
we show the critical point (CP) evolution with $\beta$ 
in NNLO (the solid line with circles) 
and NLO (the dashed line with circles).
We find that the CP stays to be the tri-critical 
point (TCP) in NNLO.
This point is different from NLO results, 
where TCP starts to deviate from 
the second order boundary at $\beta \simeq 4.5$ 
and becomes critical end point (CEP) 
for larger $\beta$~\cite{PDevol,Miura:2008gd}.
While the temperature of CP decreases in both NNLO and NLO,
we find different behavior in the chemical potential of CP, $\mu_\mathrm{CP}$.
In NNLO (NLO), CP moves in the smaller (larger) $\mu$ direction
with increasing $\beta$ in the range $\beta \gtrsim 3$.
The behavior in NNLO is favorable,
since the phase transition 
with one species of unrooted staggered fermion at $\mu=0$
is numerically shown to be the first order
in the continuum region.\cite{D'Elia:2002gd}

The above behavior of $\mu_\mathrm{CP}$ is caused
by the two temporal plaquette diagram, NNLO-TT2.
In NNLO, the diagrams with connected two plaquette configurations
shown in Fig.~\ref{Fig:NNLOdiagram}
contribute to the effective action.
In Fig.~\ref{Fig:CEP-evolution-truncation-NNLO-A},
we show the effects of each NNLO diagram on the CP evolution.
From the left, we consider the effective action terms only
from the NNLO-TT2, NNLO-SS2, NNLO-TS1+NNLO-TS2,
NNLO-TT1 or NNLO-SS1 diagram.
We find only the NNLO-TT2 shifts CP in the lower $\mu$ direction.
The other NNLO diagrams do not modify the direction of CP evolution
with $\beta$.
The temporal hopping of quarks, that is the thermal effect of quarks,
seems to be essential for the suppression of $\mu_\mathrm{CP}$
at larger $\beta$.

In Fig.~\ref{Fig:phase-diagram}, 
we display the phase diagram at $\beta=6.0$.
The NG phase ($\sigma \neq 0$) appears 
in low $T$ and low $\mu$ region,
and the Wigner phase ($\sigma=0$) appears 
at high $T$ or high $\mu$.
We find that the second-order phase boundary 
obtained by $C_2=0$ exceeds 
the first-order one in low $T$ region.
Between those boundaries,
we observe an interesting matter state
where the effective chemical potential 
$\tilmu$ is adjusted
to be around the quark excitation energy $E_q$.
In other words, the chiral symmetry
is partially restored 
to satisfy the balance relation
$\tilmu\simeq E_q$,
which leads to an intermediate value of
the quark number density $\rho_q$.
In this partially chiral restored (PCR) matter,
$N_c$ quarks are co-moving
and they have a smaller constituent quark mass;
the former shows the confined nature
and the latter is due to the partial chiral restoration.
When the PCR matter appears,
two sequential phase transitions take place
as $\mu$ becomes large with
the balance relations $\tilmu\simeq E_q(\sim\tilde{m}_q)$.
This phase transition pattern
is consistent with the implication
from the large $N_c$ argument, 
{\em i.e.} the quarkyonic transition\cite{QY_McLerran}
as discussed in our previous works.\cite{PDevol,Miura:2008gd}
In order to clarify the relation of the PCR phase and the quarkyonic matter,
it would be necessary to take account of the deconfinement transition,
which will be reported elsewhere.\cite{NNLOwithPolyakov}

While it is still uncertain
that the PCR matter
realizes in the continuum limit,
we emphasize that
this third phase (besides NG and Wigner phases)
has been commonly observed in the $T$-$\mu$-$\beta$ space
in the NLO ~\cite{PDevol,Miura:2008gd}
and NNLO SC-LQCD via the self-consistent
treatment of the vector potential $\wt$.
Compared to NLO results,\cite{PDevol,Miura:2008gd}
PCR phase appears in the lower $T$ region in the present work.
NNLO-TT1 and NNLO-TT2 are found to be responsible for such a tendency.

\section{Summary}
\label{Sec:Summary}
In this paper, we have derived 
an analytic expression of the effective potential
at finite temperature and density
including the next-to-next-to-leading (NNLO) effects
in the strong coupling expansion of the lattice QCD,
and investigated NNLO effects
on the chiral phase transition and its phase diagram.
We adopt one species of 
unrooted staggered fermion corresponding to $N_f=4$
in the continuum region.
Effective action terms have been systematically evaluated 
based on the strong coupling expansion. 
We have concentrated on the leading order of the $1/d$ expansion,
then the NNLO effective action terms are
generated from six types of two plaquette configurations.
We have applied 
the extended Hubbard-Stratonovich transformation~\cite{PDevol,Miura:2008gd}
in order to bosonize fermion interaction terms.
We encounter those terms 
containing the interaction between the next-to-nearest neighbor
sites $(x,x+2\hat{0})$ arising from $1/g^4$ terms of the effective action. 
These effects can be evaluated
by introducing the gluonic dressed fermion,
which leads to modifications of temporal quark hopping and mass terms.
We have obtained the effective potential as a function of 
temperature ($T$), chemical potential ($\mu$) 
and the two order parameters: 
the chiral condensate ($\sigma$) and the vector potential ($\wt$).
The equilibrium is determined from the stationary condition
of the effective potential with respect to the auxiliary fields.
NLO and NNLO effects result in 
modification of the wave function renormalization factor,
quark mass and chemical potential. 
The effective chemical potential reflects
the repulsive contribution of the vector potential.

We have found that
the critical temperature at $\mu=0$ ($T_{c,\mu=0}$)
is largely suppressed with increasing $\beta$.
In comparison, the chemical potential at $T=0$ ($\mu_{c,T=0}$)
is a slightly increasing function of $\beta$.
These behaviors of the critical temperature and chemical potential
in NNLO are consistent with those in NLO in the region $\beta\le 6.0$.
The behaviors $\mu_{c,T=0}$ and $T_{c,\mu=0}$
are understood from the quark mass reduction,
and its cancellation with quark chemical potential reduction,
respectively.

We have found that
the critical point (CP) moves to lower $T$ direction 
with increasing $\beta$
in both NLO and NNLO.
In NNLO, the chemical potential of the critical point 
decreases with increasing $\beta$.
This shift is opposite to the NLO case with $\beta \gtrsim 3.0$~\cite{PDevol,Miura:2008gd}.
Thus, the critical point (CP) flow with increasing $\beta$
is sensitive to $1/g^4$ effects,
and the first-order transition tends to dominate the phase boundary
as $\beta$ is increased in NNLO.
The NNLO CP flow would be favorable
because the phase transition at $\mu=0$ is shown to be
the first order for the unrooted staggered fermion in the continuum
region.\cite{D'Elia:2002gd}
We have found that
the next-to-nearest neighbor interaction makes the CP flow
go to a lower $\mu$ direction.

In low $T$ and high $\mu$ region,
we have found the partially chiral restored (PCR) phase
where effective chemical potential is always
adjusted to be around the quark excitation energy.
The PCR matter is obtained also in NLO as long as
the two order parameter ($\sigma$ and $\omega_{\tau}$) are introduced.
The appearance of PCR matter would be a common
consequence in NLO and NNLO SC-LQCD 
with finite coupling effects.

There are several points to be studied in future.
Firstly, the Polyakov loop can be constructed
by using two plaquettes in the NNLO SC-LQCD with $N_{\tau}=2$.
This effect has not been considered in the present analyses,
and quarks are confined.
It is interesting to investigate
the deconfinement transition in addition to
the chiral phase transition by considering the Polyakov loop effects.
Secondly, we should study the higher order of the $1/d$ expansion,
which contains the spatial baryon hoppings and
could play an essential role at finite density on the lattice.

\section*{Acknowledgments}

We would like to thank Professor Philippe de Forcrand for useful
discussions.
This work was supported in part
by the Grant-in-Aid for Scientific Research by MEXT and JSPS
(nos. 17070002 and 19540252),
the Yukawa International Program for Quark-hadron Sciences (YIPQS),
and by the Grant-in-Aid for the global COE program
'The Next Generation of Physics, Spun from Universality and Emergence'
from MEXT.

\appendix




\begin{thebibliography}{99}
\bibitem{Muller:2006ee}
For a recent review, see
  B.~M\"uller and J.~L.~Nagle,
  Ann.\ Rev.\ Nucl.\ Part.\ Sci.\ {\bf 56}, (2006) 93.



\bibitem{MC-sign-problem}
For a recent review, see
  S.~Muroya, A.~Nakamura, C.~Nonaka and T.~Takaishi,
  Prog.\ Theor.\ Phys.\  {\bf 110} (2003), 615;\\
%
  F.~Karsch,
  Prog.\ Theor.\ Phys.\ Suppl.\  {\bf 153} (2004), 106.


\bibitem{Wilson:1974sk}
  K.~G.~Wilson,
  Phys.\ Rev.\  D {\bf 10} (1974), 2445.

\bibitem{Creutz}
  M.~Creutz,
  Phys.\ Rev.\  D {\bf 21} (1980), 2308;
%

  M.~Creutz and K.~J.~M.~Moriarty,
  Phys.\ Rev.\  D {\bf 26} (1982), 2166.

\bibitem{Munster:1980iv}
  G.~M\"unster,
  Nucl.\ Phys.\  B {\bf 180} (1981), 23.



\bibitem{Kawamoto:1981hw}
  N.~Kawamoto and J.~Smit,
  Nucl.\ Phys.\  B {\bf 192} (1981), 100.

\bibitem{Hoek:1981uv}
  J.~Hoek, N.~Kawamoto and J.~Smit,
  Nucl.\ Phys.\  B {\bf 199} (1982), 495.


\bibitem{Smit:1980nf}
  J.~Smit,
  Nucl.\ Phys.\  B {\bf 175} (1980), 307.

\bibitem{KlubergStern:1981wz}
  H.~Kluberg-Stern, A.~Morel, O.~Napoly and B.~Petersson,
  Nucl.\ Phys.\  B {\bf 190} (1981), 504.


\bibitem{Kaplan:1992bt}
  D.~B.~Kaplan,
  Phys.\ Lett.\  B {\bf 288} (1992), 342.

\bibitem{Neuberger:1998wv}
  H.~Neuberger,
  Phys.\ Lett.\  B {\bf 427} (1998), 353.


\bibitem{Brower:1999ak}
  R.~C.~Brower and B.~Svetitsky,
  Phys.\ Rev.\  D {\bf 61} (2000), 114511.

\bibitem{Levkova:2004xw}
  L.~Levkova and R.~Mawhinney,
  Nucl.\ Phys.\ Proc.\ Suppl.\  {\bf 140} (2005), 695.

\bibitem{IchinoseNagao}
  I.~Ichinose and K.~Nagao,
  Nucl.\ Phys.\  B {\bf 577} (2000), 279;\\
  Nucl.\ Phys.\  B {\bf 596} (2001), 231.


\bibitem{XQLuo}
  P.~Ye, X.~L.~Yu, Y.~Guan and X.~Q.~Luo,
  Mod.\ Phys.\ Lett.\  A {\bf 22} (2007), 547;\\
  X.~L.~Yu and X.~Q.~Luo,
  Mod.\ Phys.\ Lett.\  A {\bf 22} (2007), 537.



\bibitem{Damgaard}
  P.~H.~Damgaard, N.~Kawamoto and K.~Shigemoto,
  Phys.\ Rev.\ Lett.\  {\bf 53} (1984), 2211;
%

  Nucl.\ Phys.\  B {\bf 264} (1986), 1.

\bibitem{DHK1985}
	P.~H.~Damgaard, D.~Hochberg and N.~Kawamoto,
	Phys.\ Lett.\ B {\bf 158} (1985), 239.



\bibitem{Faldt:1985ec}
  G.~Faldt and B.~Petersson,
  Nucl.\ Phys.\  B {\bf 265} (1986), 197.


\bibitem{Ilgenfritz:1984ff}
  E.~M.~Ilgenfritz and J.~Kripfganz,
  Z.\ Phys.\  C {\bf 29} (1985), 79.

\bibitem{Gocksch:1984yk}
  A.~Gocksch and M.~Ogilvie,
  Phys.\ Rev.\  D {\bf 31} (1985), 877.


\bibitem{Bilic}
  N.~Bilic, K.~Demeterfi and B.~Petersson,
  Nucl.\ Phys.\  B {\bf 377} (1992), 651;\\
  N.~Bilic and J.~Cleymans,
  Phys.\ Lett.\  B {\bf 355} (1995), 266.


\bibitem{Bilic:1991nv}
  N.~Bilic, F.~Karsch and K.~Redlich,
  Phys.\ Rev.\  D {\bf 45} (1992), 3228.


\bibitem{Nishida:2003uj}
  Y.~Nishida, K.~Fukushima and T.~Hatsuda,
  Phys.\ Rept.\  {\bf 398} (2004), 281.


\bibitem{Fukushima:2003vi}
  K.~Fukushima,
  Prog.\ Theor.\ Phys.\ Suppl.\  {\bf 153} (2004), 204.

\bibitem{Nishida:2003fb}
  Y.~Nishida,
  Phys.\ Rev.\  D {\bf 69} (2004), 094501.



\bibitem{Azcoiti:2003eb}
  V.~Azcoiti, G.~Di Carlo, A.~Galante and V.~Laliena,
  J. High Energy Phys. {\bf 09} (2003), 014.


\bibitem{Kawamoto:2005mq}
  N.~Kawamoto, K.~Miura, A.~Ohnishi and T.~Ohnuma,
  Phys.\ Rev.\  D {\bf 75} (2007), 014502.


\bibitem{SCL_PLoop}
  K.~Fukushima,
  Phys.\ Lett.\  B {\bf 553} (2003), 38;
  K.~Fukushima,
  Phys.\ Rev.\  D {\bf 68} (2003), 045004.


\bibitem{Kawamoto-Shigemoto-hadron-mass}
  N.~Kawamoto and K.~Shigemoto,
  Phys.\ Lett.\  B {\bf 114} (1982), 42;

  Nucl.\ Phys.\  B {\bf 237} (1984), 128.



\bibitem{KlubergStern:1982bs}
  H.~Kluberg-Stern, A.~Morel and B.~Petersson,
  Nucl.\ Phys.\  B {\bf 215} (1983), 527.
\bibitem{Jolicoeur:1983tz}
  T.~Jolicoeur, H.~Kluberg-Stern, M.~Lev, A.~Morel and B.~Petersson,
  Nucl.\ Phys.\  B {\bf 235} (1984), 455.


\bibitem{deForcrand:2009dh}
  P.~de Forcrand and M.~Fromm,
  arXiv:0907.1915 [hep-lat].

\bibitem{Karsch:1988zx}
  F.~Karsch and K.~H.~M\"utter,
  Nucl.\ Phys.\  B {\bf 313} (1989), 541.



\bibitem{PDevol}
  K.~Miura, T.~Z.~Nakano, A.~Ohnishi and N.~Kawamoto,
  Phys.\ Rev.\ D {\bf 80} (2009), 074034.


\bibitem{Miura:2008gd}
  K.~Miura, T. Z~Nakano and A.~Ohnishi,
  Prog.\ Theor.\ Phys.\  {\bf 122} (2009), 1045.


\bibitem{D'Elia:2002gd}
  M.~D'Elia and M.~P.~Lombardo,
  Phys.\ Rev.\  D {\bf 67} (2003), 014505.




\bibitem{Fodor:2001au}
  Z.~Fodor and S.~D.~Katz,
  Phys.\ Lett.\  B {\bf 534} (2002), 87.

\bibitem{Forcrand}
  P.~de Forcrand (private communication). 

\bibitem{Gottlieb:1987eg}
  S.~A.~Gottlieb, W.~Liu, D.~Toussaint, R.~L.~Renken and R.~L.~Sugar,
  Phys.\ Rev.\  D {\bf 35} (1987), 3972.

\bibitem{Gavai:1990ny}
  R.~V.~Gavai {\it et al.}  [MT(c) Collaboration],
  Phys.\ Lett.\  B {\bf 241} (1990), 567.

\bibitem{Susskind:1976jm}
  L.~Susskind,
  Phys.\ Rev.\  D {\bf 16} (1977), 3031.

\bibitem{Sharatchandra:1981si}
  H.~S.~Sharatchandra, H.~J.~Thun and P.~Weisz,
  Nucl.\ Phys.\  B {\bf 192} (1981), 205.


\bibitem{Hasenfratz:1983ba}
  P.~Hasenfratz and F.~Karsch,
  Phys.\ Lett.\  B {\bf 125} (1983), 308.



\bibitem{cumulant}
  For example, 
  R. Kubo, J. Phys. Soc. Jap. {\bf 17} (1962), 1100.

\bibitem{NNLOwithPolyakov}
  T.~Z.~Nakano, K.~Miura and A.Ohnishi, in preparation.



\bibitem{QY_McLerran}
  L.~McLerran and R.~D.~Pisarski,
  Nucl.\ Phys.\  A {\bf 796} (2007), 83;

  Y.~Hidaka, L.~D.~McLerran and R.~D.~Pisarski,
  Nucl.\ Phys.\  A {\bf 808} (2008), 117.

\bibitem{Pisarski:1983ms}
  R.~D.~Pisarski and F.~Wilczek,
  Phys.\ Rev.\  D {\bf 29} (1984), 338.

\bibitem{Kronfeld:2007ek}
  A.~S.~Kronfeld,
  PoS {\bf LAT2007} (2007), 016.
\bibitem{Sharpe:2006re}
  S.~R.~Sharpe,
  PoS {\bf LAT2006} (2006), 022.

\bibitem{Creutz:2007yg}
  M.~Creutz,
  Phys.\ Lett.\  B {\bf 649} (2007), 230.

\bibitem{convergence}
  E.~Seiler,
  Lect.\ Notes Phys.\  {\bf 159} (1982), 1;
  K.~Osterwalder and E.~Seiler,
  Annals Phys.\  {\bf 110} (1978), 440.

\bibitem{Boyd:1991fb}
  G.~Boyd, J.~Fingberg, F.~Karsch, L.~Karkkainen and B.~Petersson,
  Nucl.\ Phys.\  B {\bf 376} (1992), 199.

\bibitem{Fukushima:2003fw}
  K.~Fukushima,
  Phys.\ Lett.\  B {\bf 591} (2004), 277 .

\bibitem{Ratti:2005jh}
  C.~Ratti, M.~A.~Thaler and W.~Weise,
  Phys.\ Rev.\  D {\bf 73} (2006), 014019 .

\bibitem{Kyushu}
  K.~Kashiwa, H.~Kouno, M.~Matsuzaki and M.~Yahiro,
  Phys.\ Lett.\  B {\bf 662} (2008), 26;
%
  Y.~Sakai, K.~Kashiwa, H.~Kouno and M.~Yahiro,
  Phys.\ Rev.\  D {\bf 77} (2008), 051901.
\end{thebibliography}
\end{document}